\documentclass[5p,times]{elsarticle}

\usepackage{graphicx,miller,bm,booktabs,float,nicefrac,url,breakurl,color,multirow,microtype,upgreek,amsmath}
\usepackage{mathtools, cuted}
\usepackage[export]{adjustbox}

\usepackage{siunitx}[=v2]
\usepackage[version=4]{mhchem}
\usepackage[breaklinks,hidelinks]{hyperref}

\DeclareMathAccent{\wtilde}{\mathord}{largesymbols}{"65}

\begin{document}

\begin{frontmatter}
\title{Microscale stress-geometry interactions in an additively manufactured NiTi cardiovascular stent: A synchrotron dual imaging tomography and diffraction study }

\author[a]{Himanshu Vashishtha}
\author[b]{Parastoo Jamshidi}
\author[b]{Anastasia Vrettou}
\author[c]{Anna Kareer}
\author[c]{Michael Goode}
\author[d]{Hans Deyhle}
\author[e]{Andrew James}
\author[e]{Sharif Ahmad}
\author[e]{Christina Reinhard}
\author[b]{Moataz M. Attallah}
\author[a]{David M. Collins}
\ead{dmc51@cam.ac.uk}
\affiliation[a]{organization={Department of Materials Science \& Metallurgy, University of Cambridge},
        addressline={27 Charles Babbage Road}, 
       city={Cambridge},
       postcode={CB3~0FS}, 
       country={United~Kingdom}}
\affiliation[b]{organization={School of Metallurgy and Materials, University of Birmingham},
        addressline={Edgbaston}, 
        city={Birmingham},
        postcode={B15~2TT}, 
        country={United~Kingdom}}
\affiliation[c]{organization={Department of Materials, University of Oxford},
        addressline={Parks Road}, 
        city={Oxford},
        postcode={OX1~3PH}, 
        country={United Kingdom}}
\affiliation[d]{organization={Department of Biomedical Engineering, University of Basel},
        addressline={Petersplatz 1}, 
        city={4001 Basel},
        country={Switzerland}}
\affiliation[e]{organization={Diamond Light Source Ltd.},
        addressline={Harwell Science and Innovation Campus}, 
        city={Didcot},
        postcode={OX11~0DE}, 
        country={United~Kingdom}}

\begin{abstract}
This study explores cardiovascular stents fabricated using laser powder bed fusion (LPBF); an emerging method to offer patient-specific customisable parts. Here, the shape memory alloy NiTi, in a near equiatomic composition, was investigated to deconvolve the material response from macroscopic component effects. Specifically, stress-geometry interactions were revealed, in-situ, for a minaturised cardiovascular stent subjected to an externally applied cylindrical stress whilst acquiring synchrotron X-ray imaging and diffraction data. The approach enabled the collection of spatially resolved micromechanical deformation data; the formation of stress-induced martensite and R-phase was evident, occurring in locations near junctions between stent ligaments where stress concentrations exist. In the as-fabricated condition, hardness maps were obtained through nanoindentation, demonstrating that the localised deformation and deformation patterning is further controlled by porosity and microstructural heterogeneity. Electron backscatter diffraction (EBSD) supported these observations, showing a finer grain structure near stent junctions with higher associated lattice curvature. These features, combined with stress concentrations when loaded will initiate localised phase transformations. If the stent was subjected to repeated loading, representing in-vivo conditions, these regions would be susceptible to cyclic damage through transformation memory loss, leading to premature component failure. This study highlights the challenges that must be addressed for the post-processing treatment of LABF-processed stents for healthcare-related applications.

\end{abstract}


\end{frontmatter}

\section{Introduction}
Near equiatomic NiTi shape memory alloys have gained significant attention for biomedical applications due to their capability for absorbing large strain, generally known as superelasticity, and returning to their original shape after unloading, coined as the shape memory effect. The shape memory alloys have been effectively employed in vascular stents, orthodontic arc wires, orthopedic devices, etc.  These vascular stents are essential in restoring/controlling the blood flow through blocked or narrow arteries. Various alloys  for stent applications have been explored, including 316L stainless steel \cite{Walke2005,Kapnisis2014}, CoCr alloys \cite{Sweeney2014,Sweeney2015} Mg alloys \cite{Grogan2014,Amani2017}, and NiTi alloys \cite{Hsiao2014,McGrath2014}. From these, NiTi has emerged as a promising candidate. However, NiTi alloys may also encounter the formation of the R-phase (an intermediate rhombohedral structure and lower barrier to formation) instead of B19' phase (monoclinic structure and more thermodynamically stable than the R-phase) during the deformation of parent B2 phase (B2 cubic structure) \cite{Feng2020,Duerig2015}. The R-phase formation is considered detrimental for mechanical properties and Zhao et al. \cite{Zhao2023} stated that inhibiting R-phase formations can avoid crack formation during fatigue testing.  

The metallic stents are mainly manufactured by welding \cite{Korei2022a,Dong2018}, laser cutting \cite{Biffi2022,Fu2014}, electrospinning \cite{Cheng2022,Lee2023}, and additive manufacturing methods \cite{Li2023,Wiesent2023,Safdel2021}. Out of the existing methods, laser cutting limits the stent material and deteriorates the surface quality by forming a heat-affected zone \cite{Fu2015}. Electrospinning is not ideal for producing complex-shaped stents and is generally employed for small caliber stents or coating metal stents, limiting its use \cite{Pan2021}. Additive manufacturing has the potential to provide a single-stop solution for all secondary steps necessitated in stent production. Laser powder bed fusion (LPBF) may allow freedom of design, materials waste minimization, cost-effectiveness, and complex printable structures \cite{Korei2022b}. The LPBF technique has successfully been used for stent fabrication by utilizing the laser source and layered deposition of a metallic powder premix of the desired composition \cite{Chowdhury2022}. At present, stents are oriented to customized patient-based design, and the employment of additive manufacturing offers endless possibilities for customized stent fabrication \cite{Pan2021}. 

The stent geometry played a crucial role in defining the anticipated life of the stent and has evolved over time from the initial coil-based shape to the latest sequential rings shape (where a `Z'-shaped structure known as the strut is connected by hinges) \cite{Pan2021,Stoeckel2002,Ahadi2023}. Further, Lisa et al. \cite{Wiesent2022} performed a computational analysis on the role of geometrical irregularities over the mechanical response of 316L stainless steel stents. They stated that the geometrical irregularities encourage the higher localized strains and consequently the stent fracture. Similarly, other studies undertook computational analysis \cite{Sweeney2014,Berti2021,Berti2019,Jedwab1993,RAGHUNATHAN20081059} to predict the preferred fracture locations in the stents. Sweeney et al. \cite{Sweeney2014} modelled the fatigue response of a CoCr stent, predicting the high probability of failure points along the inner radius of the top strut or close to the tri-junction of the stent. They rationalized the high probability of failure points by arguing there is a notch point with an associated high-stress concentration. Some researchers have measured the micromechanical response of the stents \cite{Safdel2023,Jamshidi2022}. However, a detailed in-situ experimental investigation at the sub and near crystal length scale is lacking. 

Synchrotron X-ray diffraction (SXRD) has emerged as the most powerful tool to perform in-situ  characterisation of materials. The SXRD offers a wide range of advantages over laboratory-based XRD in terms of tunability of wavelengths, increased sensitivity, high spectral resolution, and fast scan, suitable for a wide range of materials and sample thicknesses. There are numerous prior studies that have applied the SXRD techniques to the NiTi alloy \cite{Feng2020,Bian2022,Sedmak2015,Urbina2017,Yu2016}. These experiments are commonly used to measure crystal structure characteristics for all crystalline phases present, including lattice parameters, that can be related to crystal orientation dependent lattice strains and texture development. This study represents the first attempt to investigate the in-situ deformation behaviour of stents in their actual component form. Moreover, X-ray synchrotron imaging coupled with diffraction increase the effectiveness by revealing surface features and volumetric alterations \cite{Reinhard2021}; such coupled methods are particularly exciting as phenomena at the microscopic and macroscopic length scales can be directly related.

In this first-of-its-kind investigation, miniaturised cardiovascular stents were additively manufactured by the LPBF method, with the dimensions selected to match experimental constraints. Stents were  characterised, in-situ, by high-energy synchrotron X-ray dual imaging and diffraction at the DIAD beamline, Diamond Light Source. The stents were measured in three states; the as-printed state to observe the initial stresses, under deformation to observe any high-stress points and possible failure locations, and in the unloaded state to quantify the retained stresses. Moreover, the transformation temperatures were recorded to verify the applicability of the stents. Additionally, the nanoindentation and electron backscatter diffraction (EBSD) were employed to assess the crystallographic, microstructure and sub-crystal characteristics. In conjunction with the spatially-resolved deformation behaviour, this study seeks to understand at the micro-scale, the characteristic features of the material and the geometry, to assess the feasibility and future challenges for LPBF processed cardiovascular stents. 

\begin{figure}[h!]
\centering
    \includegraphics[width=85mm]{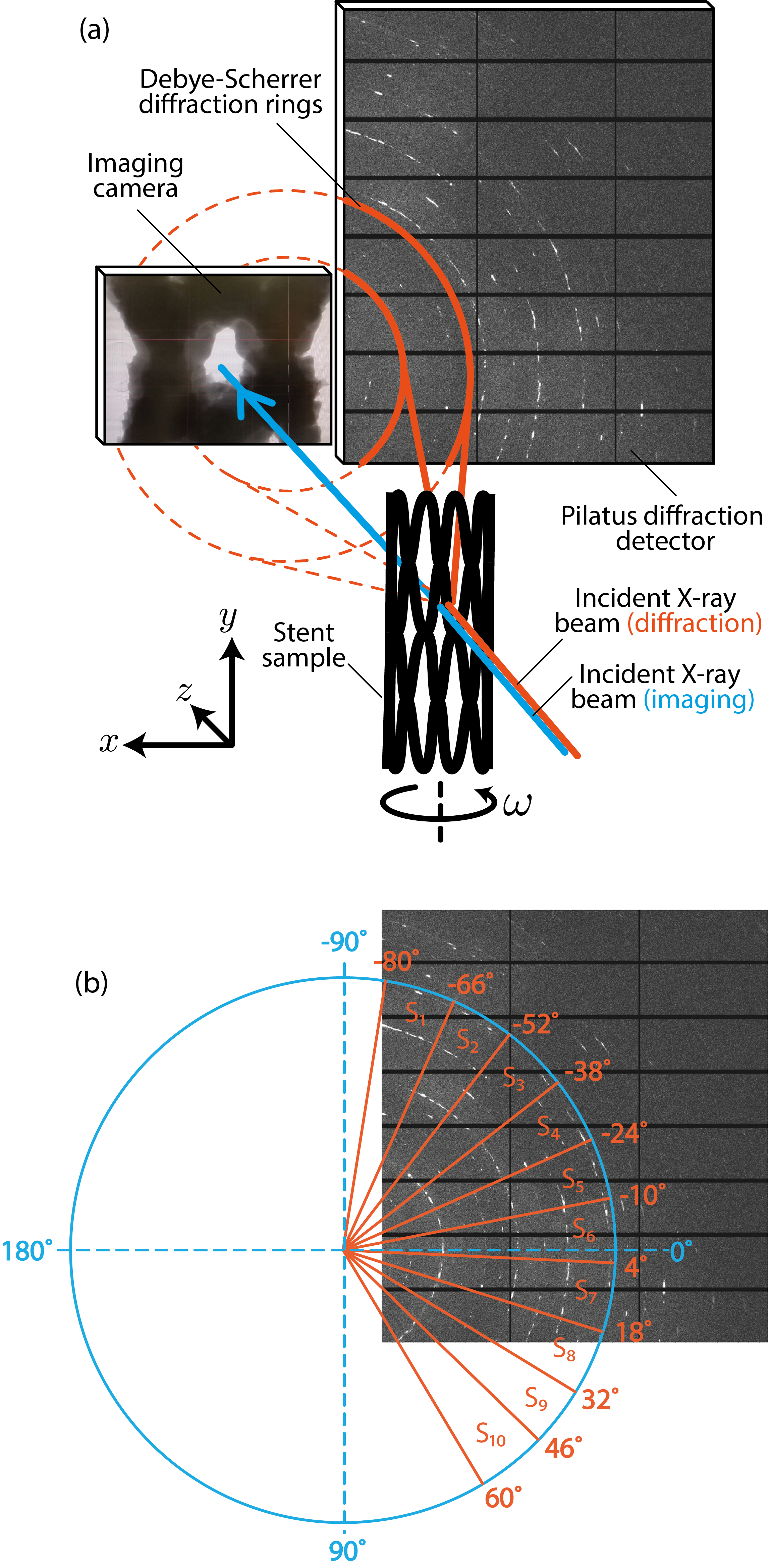}
    \caption{Experimental setup on the DIAD beamline, Diamond Light Source. The configuration is set up for the quasi-simultaneous collection of diffraction and imaging data; here applied to a mm-scale nitinol stent.}
    \label{fig:Setup}
\end{figure}

\section{Experimental Method}
\subsection{Material}
In this work, pre-alloyed equiatomic NiTi powders prepared by plasma wire atomization process were supplied by Advanced Powders and Coatings (AP\&C). A Coulter LS230 laser diffraction particle size analyser was used to measure the particle size. The average particle size, D50, was $\sim$32.2\,$\upmu$m with a spherical morphology with  some smaller satellite particles; this indicates good flowability for subsequent processing. Zigzag-shaped stents were additively manufactured using the LPBF method. A contour scanning strategy was followed to achieve the fine stent geometry. A Yb-fibre Concept Laser M2 cutting system, under an argon atmosphere $<1000$\,ppm, was used as a laser power source. The laser parameters were fixed at a laser power of 80\,W, with a contour scan at a speed of 250\,mm/s, a 60$\upmu$m, spot size, 20\,$\upmu$m layer thickness.  After fabrication, the stents were cleaned in a chemical mixture of HF:HNO$_3$:H$_2$O (1:2:3) for twelve minutes and then cleaned with distilled water and ethanol \cite{Jamshidi2022}. The stent length was 3.81\,mm with an outer diameter of 1.45\,mm. These dimensions were tailored to suit the probed X-ray volume accessible via synchrotron methods, as described later.

\subsection{Microstructural characterization }

To study the microstructure, the samples were mounted in cold-setting powder epoxy then metallographically prepared using abrasive media. This comprised successive grinding with SiC paper (200, 400, 600, 800, 1200, 2500, and 4000 grit size), finishing with cloth polishing in a 0.05\,$\upmu$m suspended silica solution. Etching was performed in an etchant containing HF (3.2\%), HNO$_3$ (14.1\%), and H$_2$O (82.7\%), for 20 seconds \cite{Saedi2018}. The laser scanning trajectories and microstructural features were observed using a field emission scanning electron microscope (FESEM, JEOL 7000, Japan). The stents were also characterised using electron backscattered diffraction (EBSD) analysis, here prepared by electrochemical polishing in 790\,ml acetic acid and 210\,ml perchloric acid solution at 10\,V for 30\,s \cite{Vashishtha2022}. The EBSD scans (Nordlys detector, Oxford Instruments, U.K.) were acquired at two magnifications, one with a step size of 1 \,$\upmu$m, and another at 0.3\,$\upmu$m. Each were obtained with a 30\,kV operating voltage. 

\subsection{Nanomechanical testing}

Hardness mapping was performed on the as-fabricated condition of the stent. A load-controlled G200 nanoindenter (KLA Tencor, U.S.A.) with a Berkovich indenter tip at a 2.2\,mN load and a $<0.01$\,nm displacement resolution was used. The indent spacing was fixed at 1$\upmu$m, approximately 8 times the indentation depth, and a grid mapping protocol was employed to spatially describe the stent hardness. 

\subsection{Synchrotron diffraction and imaging}
A synchrotron X-ray experiment was performed at the dual imaging and diffraction (DIAD) beamline, Diamond Light Source, U.K.  Figure \ref{fig:Setup} illustrates the experimental setup, including the incident and measured X-ray beam paths. It also shows the quasi-simultaneous data collection arrangements for X-ray diffraction and X-ray computed tomography (CT). A monochromatic incident X-ray beam, operating at 30\,keV, calibrated with a CeO$_2$ standard, was used for all acquired X-ray measurements in a transmission geometry. Following alignment of the cylinder axis stent onto the rotation centre of the sample stage, a mapping strategy was adopted to obtained spatially resolved detail of the stent. In an imaging configuration, the sample was rotated, about $\omega$ (see Fig. \ref{fig:Setup}, whilst acquiring radiographs at steps of 0.05$^\circ$ over an rotational range of 180$^\circ$ with an acquisition time of 0.05 seconds and a $1.28\times1.08$ mm$^2$ field of view. This was captured on a PCO.edge 5.5 CMOS detector, providing an effective pixel size of $0.5\times0.5$\,$\upmu$m$^2$. 

Diffraction data was also recorded, here using a Dectris PILATUS3 X CdTe 2M detector, with a pixel size of $172\times172$\,$\upmu$m$^2$, which was situated at a distance of 0.365\,m from the sample. Using a beam-size of $25\times25$\,$\upmu$m$^2$, diffraction patterns with an acquisition time of 1\,second  were acquired at $0.5^\circ$ increments over a $360^\circ$ rotation about $\omega$. This was repeated for 31 positions along the stent cylinder axis ($y$), covering a total of 0.75\,mm.

\begin{figure}[h!]
\centering
    \includegraphics[width=85mm]{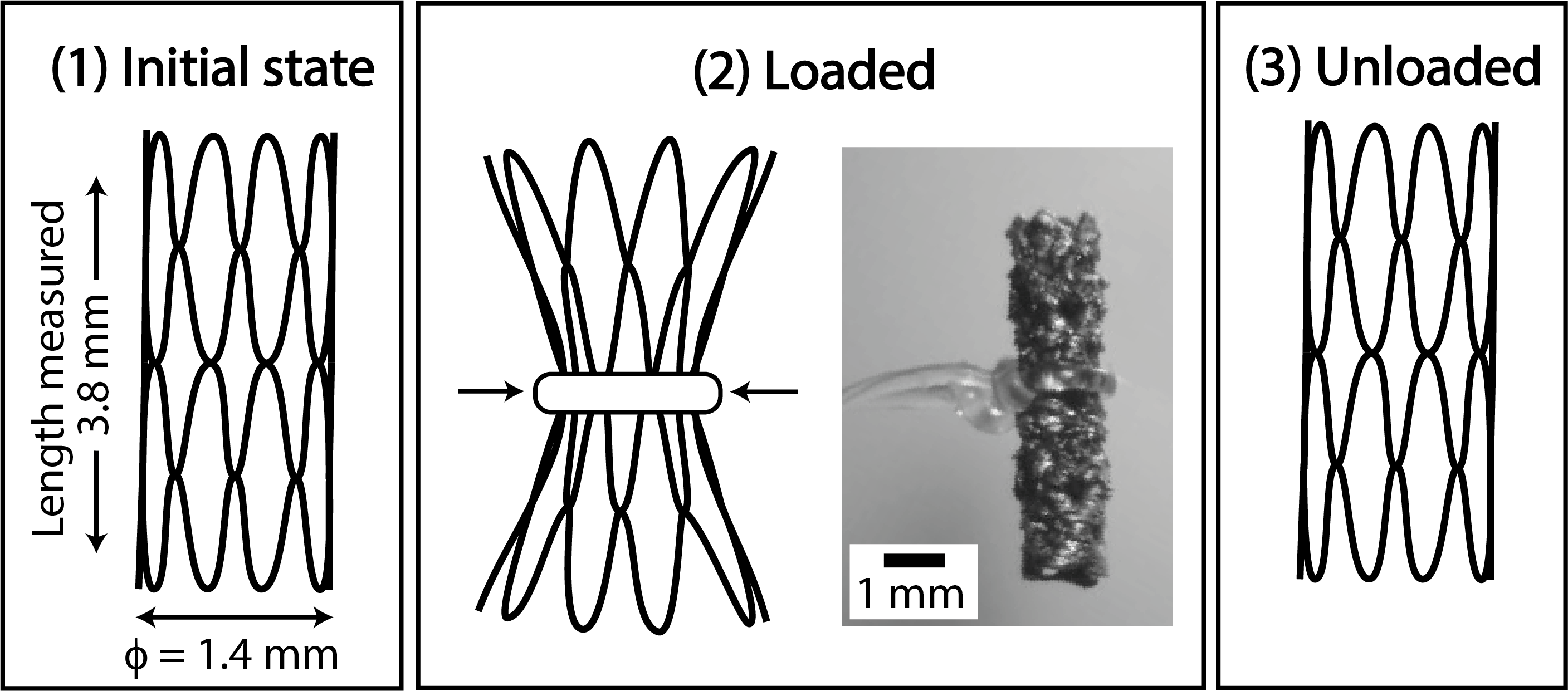}
    \caption{Three deformation conditions were studied on the same sample: (1) the initial as fabricated state, (2) loaded with a compressive hoop stress, and finally (3) after unloading. }
    \label{fig:DeformationSchematic}
\end{figure}

The data collection strategy was repeated on the stent for several deformation states; this is illustrated in Figure \ref{fig:DeformationSchematic}. The first diffraction and tomography experiment was performed in the zero load (as fabricated) condition. A compressive hoop stress was applied by squeezing the stent with nylon wire, as shown in Fig. \ref{fig:DeformationSchematic} (b). Measurements were then repeated after unloading. In each deformation condition, data were acquired from the same stent region.

\subsubsection{Data Fitting}

All diffraction patterns were stored at image files, then radially integrated into azimuthal ($psi$) sectors. The data was integrated into 10 sectors of equivalent size over the azimuthal range of the Debye Scherrer patterns captured by the detector; $-80\leq\psi\leq60$, as shown in Fig. \ref{fig:Setup} (b). Individual reflection fitting using a Pseudo-Voigt function was performed on each diffraction spectra, using an in-house written Matlab method; the reflection position and intensities were used for subsequent analysis. The lattice parameters for the B2 cubic austenite in the strain-free state were, $a_0 = 3.015$\,\AA, $\alpha = \beta = \gamma = 90^\circ$ \cite{Otsuka2005}. Lattice strains, $\varepsilon_{hkl}$ for the plane Miller indices $hkl$ were calculated using:

\begin{equation}
\varepsilon^{hkl}_{\psi} = \frac{d^{hkl}_{\psi}-d_0^{hkl}}{d^{hkl}_0}
\end{equation}

\noindent where $d^{hkl}_{\psi}$ is the d-spacing at a given azimuthal angle $\psi$ and $d_0^{hkl}$ is the reference d-spacing in the strain free state, where $d_0 = a_0/N^{1/2}$ and $N=h^2+k^2+l^2$. Reflections up to $N=10$ were fitted. For the minority phases, martensite and the R phase, only the $(2\,0\,0)_{\rm B19'}$ and $(1\,2\,2)_{\rm R}$ peaks were fitted, respectively, as their intensities were observed to be the largest, and therefore most reliable for quantification.

\subsubsection{Macromechanical Stress Determination}

Following a similar treatment of macromechanical stress variation, shown elsewhere \cite{CHEN2021116777}, the stress was calculated from the lattice strain data obtained from the austenitic B2 phase, as a function of position on the stent surface.

For a cylindrical coordinate system, $(r, \omega, z)$, the micromechanical strain in the $\omega-z$ plane, following \cite{He}, is: 

\begin{equation}
\varepsilon^{hkl}_{\psi} = p^{hkl}_{\omega\omega,\psi}\sigma_{\omega\omega} + p^{hkl}_{\omega z,\psi}\sigma_{\omega z} + p^{hkl}_{zz,\psi}\sigma_{zz}
\label{micro_strain}
\end{equation}

\noindent where $p^{hkl}_{ij,\psi}$ are plane specific stress factors and $\sigma_{ij}$ are macromechanical stress tensor components. The values of $p^{hkl}_{ij,\psi}$ can be related to the X-ray elastic constants, $S_1^{hkl}$ and $\frac{1}{2}S_2^{hkl}$ as follows \cite{Hauk}:

\begin{equation}
p^{hkl}_{\omega\omega,\psi} = S_1^{hkl} + \frac{1}{2}S_2^{hkl}\cos^2{\psi}
\label{Eq3}
\end{equation}

\begin{equation}
p^{hkl}_{\omega z,\psi} = \frac{1}{2}S_2^{hkl}\sin(2\psi)
\label{Eq4}
\end{equation}

\begin{equation}
p^{hkl}_{zz,\psi} = S_1^{hkl} + \frac{1}{2}S_2^{hkl}\sin^2(\psi)
\label{Eq5}
\end{equation}

The values of $S_1^{hkl}$ and $\frac{1}{2}S_2^{hkl}$ were calculated using the ISODEC software \cite{isodec}. This requires elastic constants, taken as $c_{11} = 162$\,GPa, $c_{12}=132$\,GPa and $c_{44}=36$\,GPa for B2 austenite in NiTi \cite{Mercier}.

The in-plane stress components $\sigma_{\omega \omega}$, $\sigma_{\omega z}$ \& $\sigma_{zz}$ can be obtained by using the lattice strains in Equation \ref{micro_strain}, here using the B2 diffraction reflections up to $N=10$, and for each of the 10 azimuthal sectors ($\bar{\psi_1} = -73^\circ$, $\bar{\psi_2} = -59^\circ$, $\bar{\psi_3} = -45^\circ$,...,$\bar{\psi_{10}} = -53^\circ$). Using these values, the macromechanical stress tensor was calculated using:

\begin{equation}
    \{\boldsymbol{\varepsilon}\} = \{\boldsymbol{\sigma}\}[\boldsymbol{p}]
    \label{Eq6a}
\end{equation}

\begin{equation}
    \{\boldsymbol{\varepsilon}\} = \{\varepsilon^{110}_{\bar{\psi_1}}~~~\varepsilon^{200}_{\bar{\psi_1}}~~~\varepsilon^{211}_{\bar{\psi_1}}~~~\varepsilon^{220}_{\bar{\psi_1}}~~~\varepsilon^{310}_{\bar{\psi_1}}~~~\varepsilon^{110}_{\bar{\psi_2}}~~~...~~~\varepsilon^{310}_{\bar{\psi_10}}\}
\end{equation}

\begin{equation}
    \{\boldsymbol{\sigma}\} = \{\sigma_{\omega\omega}~~~\sigma_{\omega z}~~~\sigma_{zz} \}
\end{equation}

\begin{strip}
\begin{equation}
    [\bf{p}] = 
\begin{bmatrix}
    p_{\omega\omega,\bar{\psi_1}}^{110} & p_{\omega\omega,\bar{\psi_1}}^{200} & p_{\omega\omega,\bar{\psi_1}}^{211} & 
    p_{\omega\omega,\bar{\psi_1}}^{220} &
    p_{\omega\omega,\bar{\psi_1}}^{310} &
    p_{\omega\omega,\bar{\psi_2}}^{110} &\dots  & p_{\omega\omega,\bar{\psi_{10}}}^{220} \\
        p_{\omega z,\bar{\psi_1}}^{110} & p_{\omega z,\bar{\psi_1}}^{200} & p_{\omega z,\bar{\psi_1}}^{211} & 
    p_{\omega z,\bar{\psi_1}}^{220} &
    p_{\omega z,\bar{\psi_1}}^{310} &
    p_{\omega z,\bar{\psi_2}}^{110} &\dots  & p_{\omega z,\bar{\psi_{10}}}^{220} \\
    p_{zz,\bar{\psi_1}}^{110} & p_{zz,\bar{\psi_1}}^{200} & p_{zz,\bar{\psi_1}}^{211} & 
    p_{zz,\bar{\psi_1}}^{220} &
    p_{zz,\bar{\psi_1}}^{310} &
    p_{zz,\bar{\psi_2}}^{110} &\dots  & p_{zz,\bar{\psi_{10}}}^{220} \\
\end{bmatrix}
\end{equation}
\end{strip}

The values of $p_{ij,\psi}^{hkl}$ were calculated using Equations \ref{Eq3}, \ref{Eq4} and \ref{Eq5}. The macromehanical stresses were finally obtained by rearranging Eq. \ref{Eq6a}:

\begin{equation}
 \{\boldsymbol{\sigma}\} = \{\boldsymbol{\varepsilon}\}[\boldsymbol{p}^+]
\end{equation}

\noindent where $[\bf{p}]^+$ is the Moore-Penrose pseudo inverse of the matrix, $[\bf{p}]$. Once the in-plane stress components were obtained, the 2D Von-Mises Stress, $\sigma_{VM}$ was calculated:

\begin{equation}
    \sigma_{VM} = \sqrt{\sigma_{\omega\omega}^2 + \sigma_{zz}^2 + \sigma_{\omega\omega}\sigma_{zz} + 3\sigma_{\omega z}^2 }
\end{equation}

For further details on the validity of the conversion between $\varepsilon^{hkl}$ and $\sigma_{ij}$, the reader is referred elsewhere \cite{stress_review}.

\subsubsection{Treatment of phase spatial distributions}
To assess the spatial distribution of the austenite, R and martensite phases, the integrated intensities of diffraction reflections associated with each were used. As a qualitative indicator of the phase proportions, the sum of the 
$(1\,1\,0)_{\rm B2}$, $(2\,0\,0)_{\rm B19'}$ and $(1\,2\,2)_{\rm R}$ were normalised to one for each location measured from the stent. Thus, a relative variation of these phases could be assessed.

\section{Results}
\subsection{Microstructure}

\begin{figure}
\centering
    \includegraphics[width=85mm]{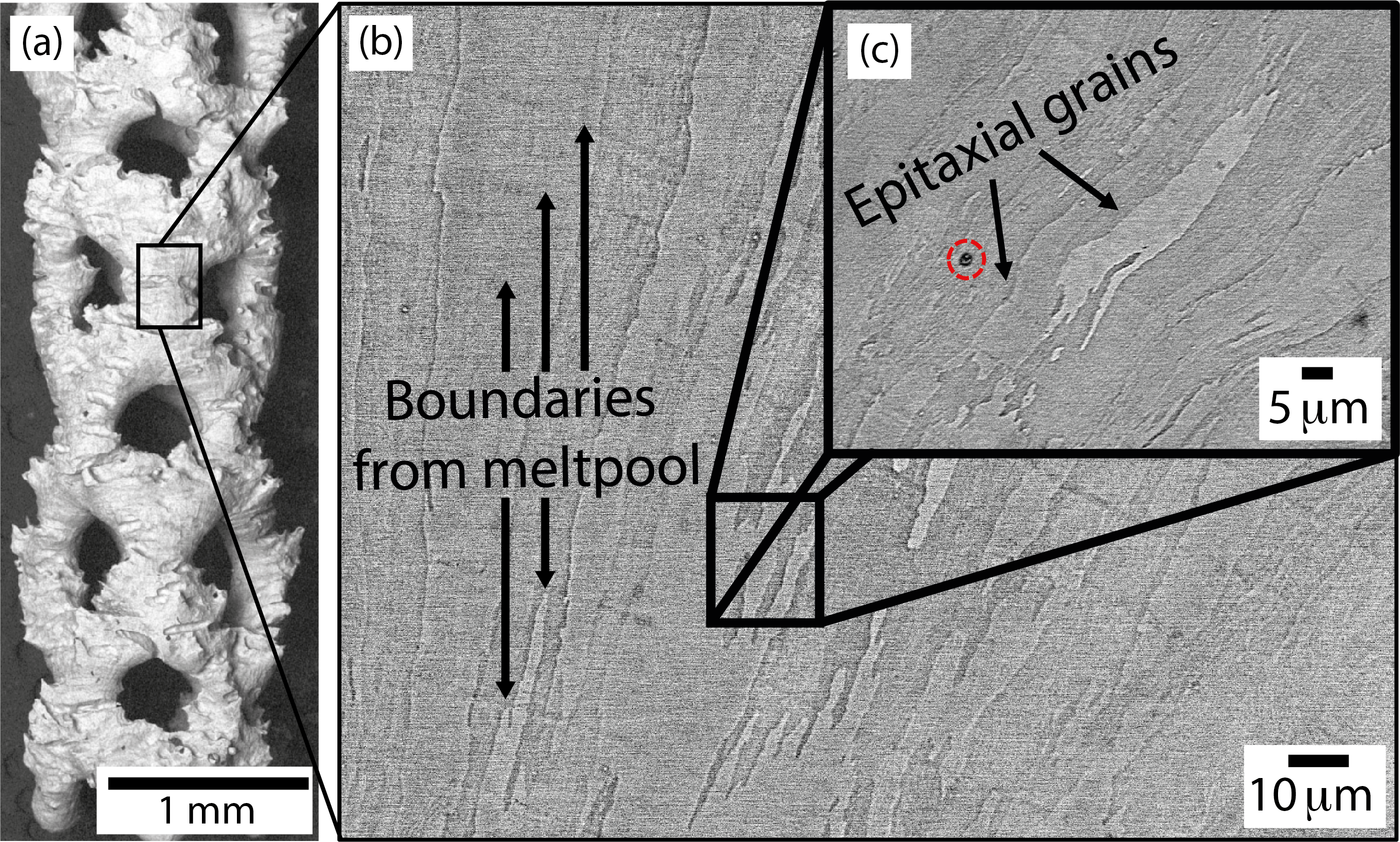}
    \caption{(a) Macroscopic view of the stent, (b) SEM image reveals surface morphology and laser tracks in the stent, and (c) depicts a higher magnification view. }
    \label{fig:InitialCondition}
\end{figure}

A contour scanning strategy was adopted during the  zig-zag-shaped stent fabricating. The complete stent was visualized under SEM and the macroscopic view was captured, as shown in Fig. \ref{fig:InitialCondition} (a). The mounted, mechanically polished, and chemically etched stent was examined under SEM; the microstructure is shown in Figure \ref{fig:InitialCondition} (b) and a magnified view is shown in the subset, Fig. \ref{fig:InitialCondition} (c). Columnar grains are seen that have grown epitaxially between the layers. Orthogonal to the grains, faint boundaries can be seen which correspond to the location of the meltpool during the manufacture, bordering the laser tracks. Some porosity is evident,  highlighted by the red circle in the subset region. These features of the microstructure are representative of all locations across the stent cross-section.

\subsection{EBSD Characterisation}

\begin{figure*}[h!]
\centering
    \includegraphics[width=170mm]{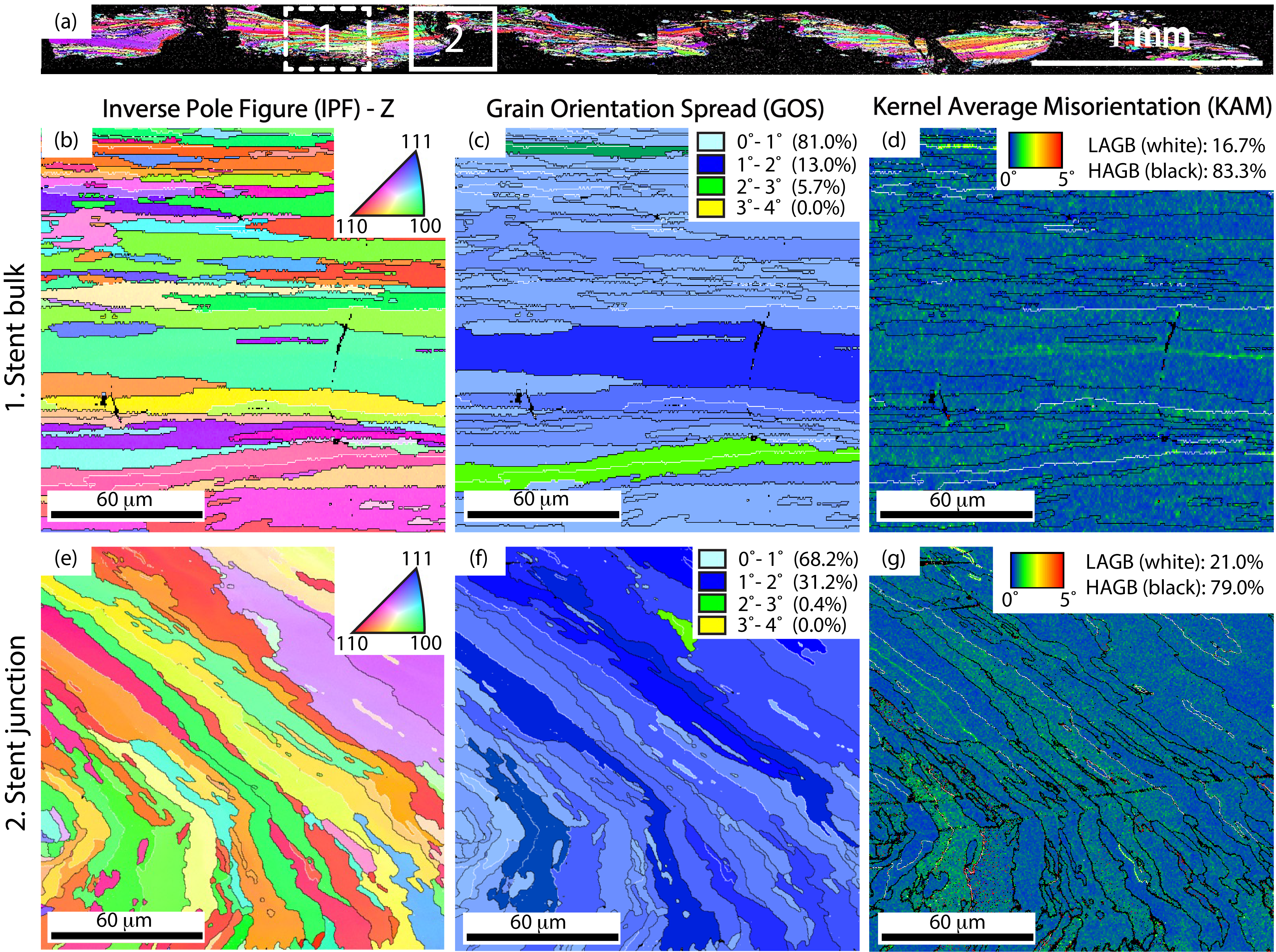}
    \caption{EBSD analysis of the stent, indexed as the B2 austenite phase. The datasets are presented as (a) an IPF-Z map of the whole stent; high-resolution IPF-Z, GOS and KAM maps of the bulk sample, within ligaments away from any junctions (b), (c) \& (d). Equivalent maps are also shown for a near junction region (e), (f) \& (g). }
    \label{fig:EBSD}
\end{figure*}

EBSD maps of the B2 austenite phase for the full stent cross-section, along the longitudinal direction as illustrated in Fig. \ref{fig:EBSD} (a), and at higher magnification in selected regions, Fig. \ref{fig:EBSD} (b) – (g). The design and geometry of the stent plays a decisive role in the location specific microstructure, which must heavily influence the functional and mechanical properties. In particular, differences are observed in the bulk of the ligaments, and to locations at the ligament junctions. Representative regions for each were acquired, marked as `1', and `2' in Fig. \ref{fig:EBSD} (a), respectively. The IPF-$Z$ maps (inverse pole figure, the colours respect to the out of plane direction) suggest that the crystallographic orientation in the bulk, Fig. \ref{fig:EBSD} (b), and the junction, Fig. \ref{fig:EBSD} (e), are similar. The grain orientation spread (GOS) is calculated as the average difference between the orientation of each pixel inside a grain and the grain average orientation. All pixels within a grain are assigned the same colour. It is reported that the deformed (higher-strain) grains have higher GOS than the recrystallized (lower-strain) grains \cite{Suresh2012}. In this investigation, the threshold GOS value is set at 4$^\circ$ and bifurcated into 4 equal ranges to identify the localised variation in bulk and junction regions. The bulk region in Fig. \ref{fig:EBSD} (f) shows higher GOS, likely due to higher levels of plasticity that result in this measured per-grain distortion, compared to the bulk region in Fig. \ref{fig:EBSD} (c). Similarly, the Kernel average misorientation (KAM) maps in Fig. \ref{fig:EBSD} (d) and (g) also validate the claim that the junction region stores higher lattice rotation (dislocations). It is considered as the dislocations are favourably formed at the low-angle grain boundaries (LAGBs). The junction region possesses 21\% of LAGBs whereas, the bulk region possesses only 16.7\%. This also validates the GOS and KAM results that the junction region accumulates more LAGB, and therefore greater dislocation density.

\begin{figure}[h!]
 \centering
    \includegraphics[width=85mm]{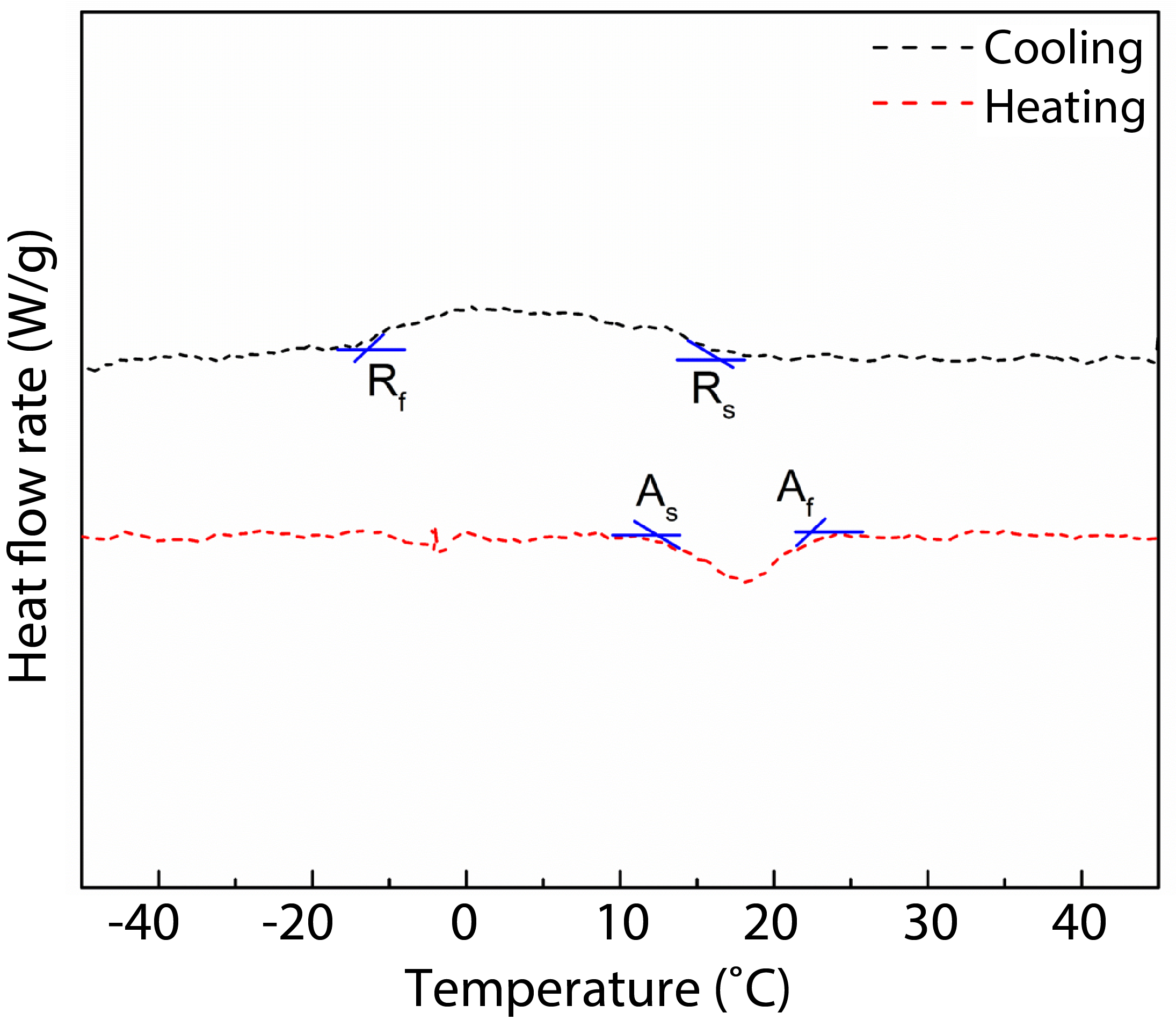}
    \caption{The DSC curve illustrating the austenitic and R-phase, start and finish temperatures. }
    \label{fig:DSC}
\end{figure}

\subsection{Transformation temperatures}
The transformation temperature of the stent was captured by a differential scanning calorimetry (DSC) test, as illustrated in Fig. \ref{fig:DSC}. Transformation peaks are not observed to be sharp for the as-printed stent. The austenitic start (A$_{\rm s}$) and austenitic finish (A$_{\rm f}$) temperatures are recorded as 22$^\circ$C and 47\,$^\circ$C whereas, the R-phase start (R$_{\rm s}$) and R-phase finish (R$_{\rm f}$) are measured as 32$^\circ$C and -14$^\circ$C. It is known that the transformation peaks sharpen when the alloy is heat treated \cite{Jamshidi2022}.

 \begin{figure*}[h!]
 \centering
    \includegraphics[width=170mm]{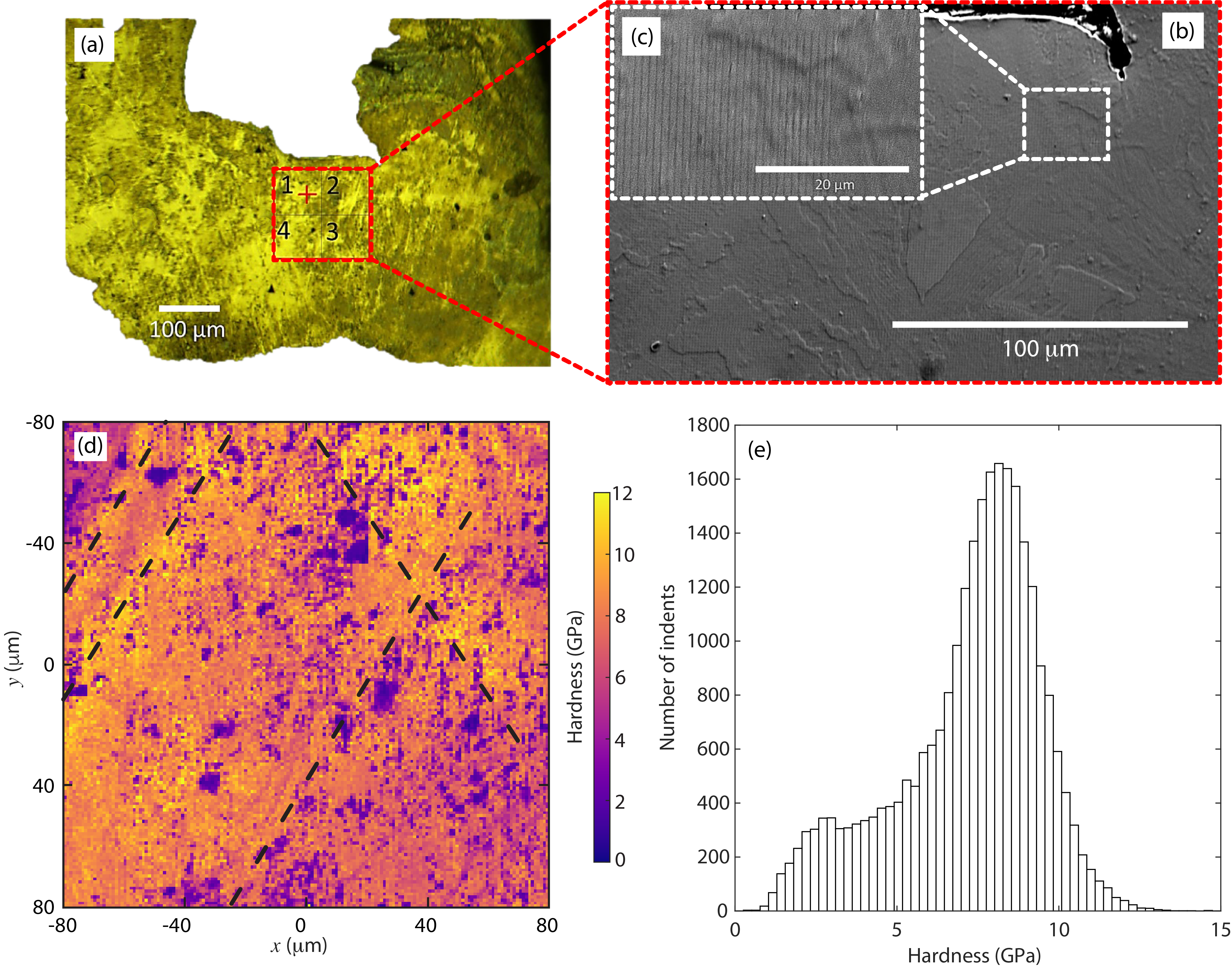}
    \caption{(a) Optical micrograph showing the indented region, (b) post-indentation SEM image, and (c) higher magnification SEM image in the inset showing the arrays of indents. (d) Hardness map measured with nanoindentation; the dashed lines are shown parallel to bands of approximately uniform hardness, and (e) corresponding histogram of hardness data.}
    \label{fig:Nanoindentation}
\end{figure*}

\subsection{Hardness mapping}

Hardness mapping of the stent was performed by nanoindentation, as shown in Fig. \ref{fig:Nanoindentation}. Fig. \ref{fig:Nanoindentation} (a) shows an optical micrograph of the indent region of $160\times 160$\,$\upmu$m$^2$ marked as a square box. The region is deemed to be sufficiently large to assess any interaction between the microhardness properties and geometry, close to a junction. Notably, this matches the high-magnification EBSD mapping dimensions, so the results probe $\sim$100 grains.  Post-indentation, the surface was captured under SEM and shown in Fig. \ref{fig:Nanoindentation} (b) and further higher magnification images revealed the arrays of indents as shown in the inset image in Fig. \ref{fig:Nanoindentation} (c).  The mapped hardness values are shown in Fig. \ref{fig:Nanoindentation} (d); several features of note can be seen. Isolated regions below 2\,GPa in hardness and $<5\,\upmu$m in diameter are pores within the material. There are bands of approximately uniform hardness, as annotated with the dashed lines, which run towards the free edge of the stent. This patterning is unsurprising given the location dependent morphology of the grain structure, shown in the EBSD results. A histogram of all measured hardness values is shown in Fig. \ref{fig:Nanoindentation} (e), with a mean hardness of $7.28\pm2.25$ GPa. There are evidently multiple populations of hardness, with one centred at $\sim3$\,GPa and another centred at $\sim3$\,GPa; the hardness is highly heterogeneous but is within the range observed in other studies \cite{Pfetzing2013}. The heterogeneity is attributed to the porosity defects (surface or sub-surface) \cite{Anton2020} and anisotropic microstructures \cite{Kok2018} associated with additive manufacturing. 


\subsection{X-ray tomography}

\begin{figure}[h!]
\centering
    \includegraphics[width=60mm]{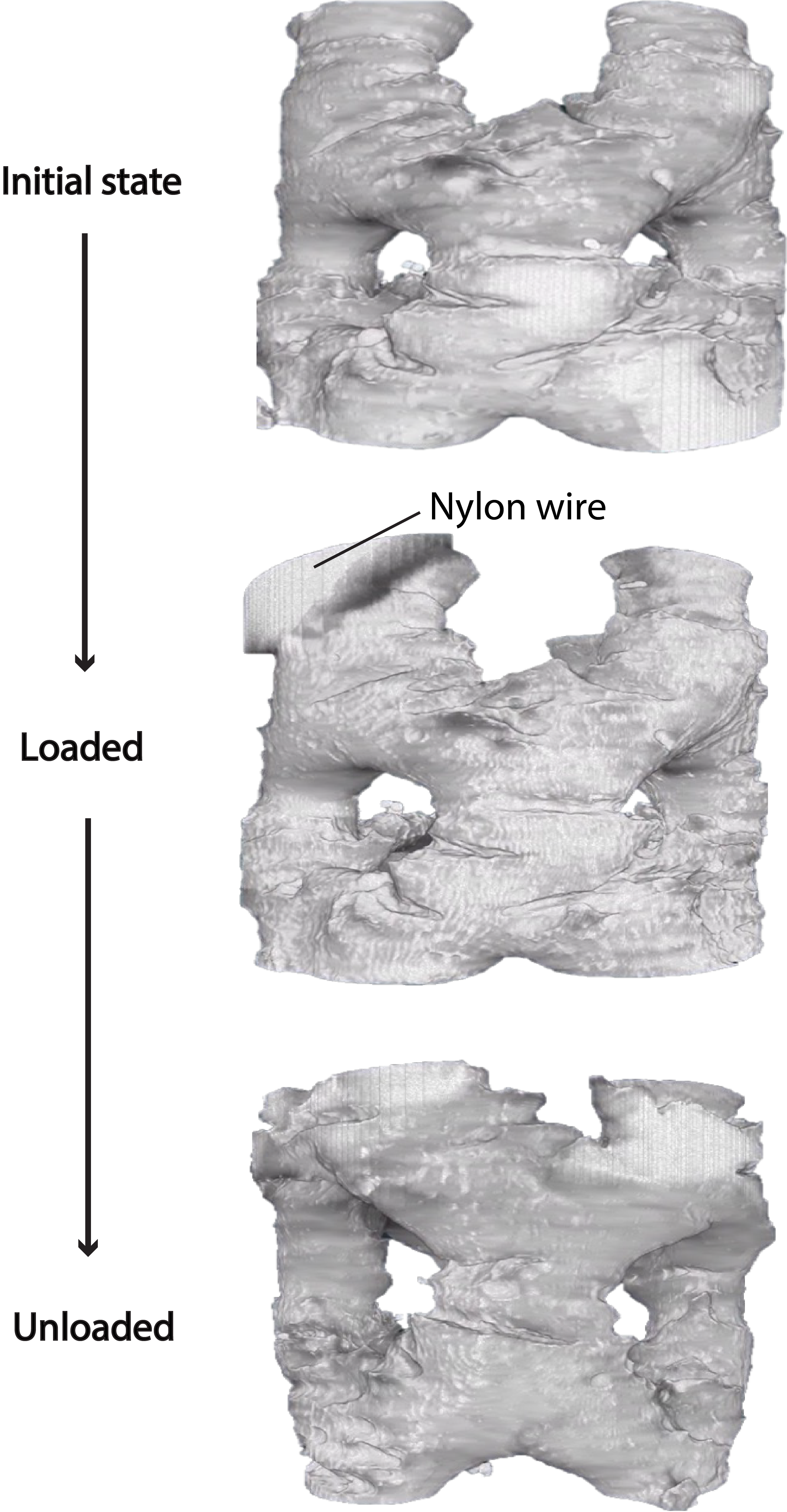}
    \caption{X-ray tomographic reconstruction of the stent in the initial, deformed and unloaded states.}
    \label{fig:Tomo}
\end{figure}

Tomographic reconstruction of the stents in all three states was performed by the commercially available software Avizo. Volumetric rendering was performed to visualize the data in a 3-dimensional space as shown in Fig. \ref{fig:Tomo}. The X-ray tomography reveals that the stent had rough or uneven surfaces;  this is commonly associated with the powder-based additive manufacturing method of stent fabrication \cite{Jacob2020}. The reconstructions were also used to estimate the macroscopic strain in the deformed condition; the inner diameter of the stent was measured, in the position of the wire, in the initial and deformed state. The circumferential strain, $\varepsilon_\omega$, was estimated to be 0.011.

\subsection{Synchrotron Diffraction}

\begin{figure}[h!]
 \centering
    \includegraphics[width=85mm]{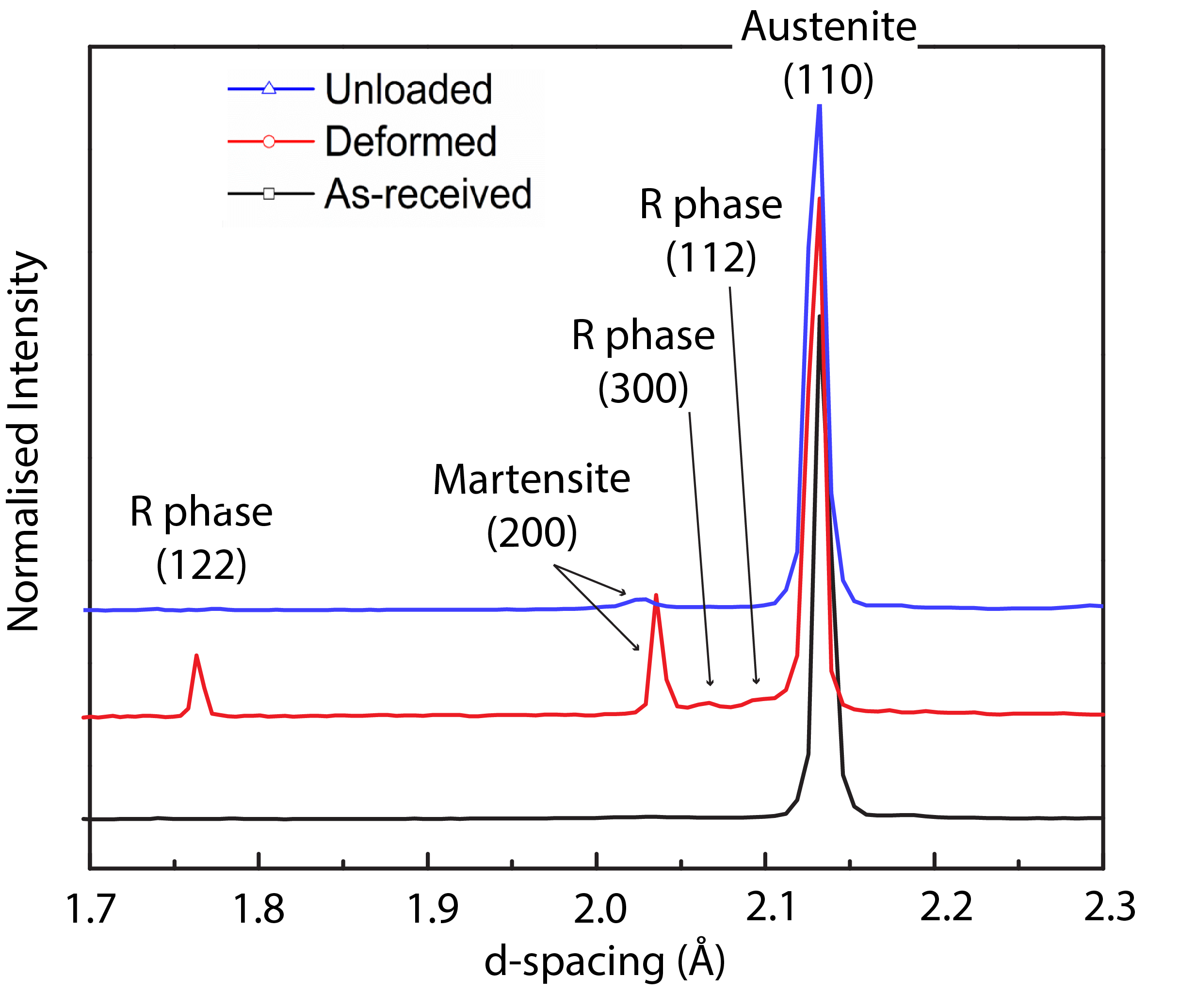}
    \caption{Representative X-ray diffraction patterns, selection from a region close to a junction. }
    \label{fig:XRD}
\end{figure}

Example indexed diffraction patterns are shown in Fig. \ref{fig:XRD}, in each of the initial, loaded and unloaded states. The stress-induced martensitic transformation was recorded for the deformed specimen; the martensitic peak of $(2 0 0)$ reflection was notably strong. After unloading, some of the martensite was retained in the specimen. Additionally, several R-phase reflections were only observed in the deformation state. The peaks were weak, indicating the phase was not present in significant volume fractions, but was unambiguously present when load was applied.

\begin{figure*}[h!]
 \centering
    \includegraphics[width=180mm]{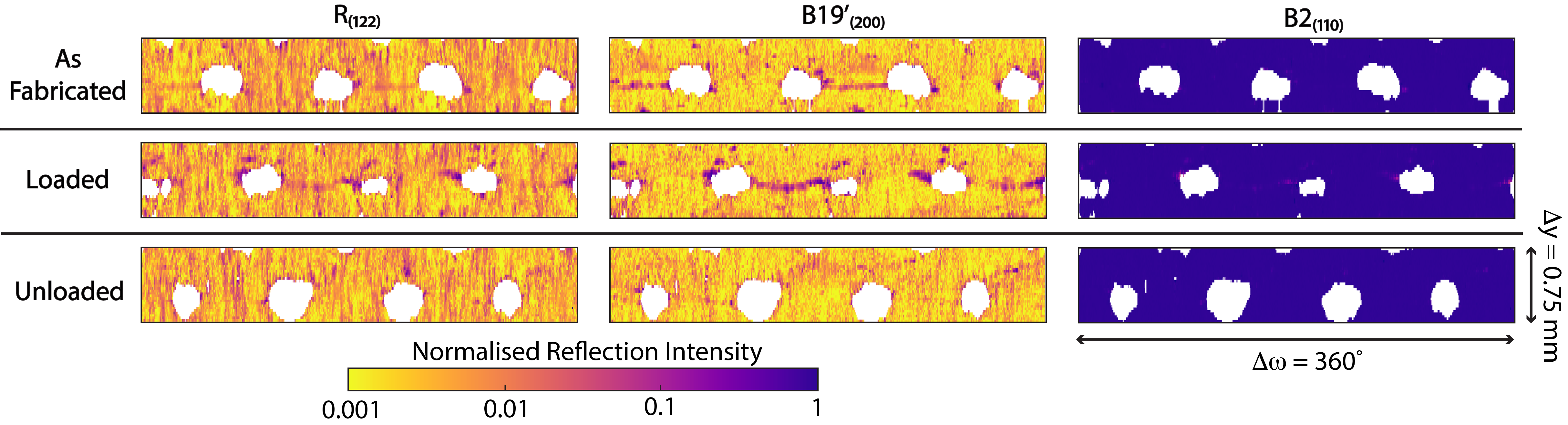}
    \caption{Normalised intensities for $R_{(122)}$ phase $B19'_{(200)}$ martensite phase and for the $B2_{(110)}$ phase, respectively for the as-fabricated, deformed and unloaded states of the stent.}
    \label{fig:Intensities}
\end{figure*}

The independent normalised intensity plots for the B2 austenite phase, B19' martensite phase, and R-phase in all three, as-received, deformed, and unloaded states are illustrated in Fig. \ref{fig:Intensities}. The empty spaces between the stent ligaments are shown here as white regions. Notably, these spaces were observed to be circular in the as-received state (Fig. \ref{fig:Intensities}) and slightly elliptical during the deformation . The circular shape was recovered after load removal. Moreover, the presence of the B19' martensite phase increases in the deformed state  compared to the as-received state, and surprisingly R-phase was also formed in the loaded state. Here, narrow bands of martensite and R-phase were evident, connecting the regions of empty space. These have followed the hoop direction ($\omega$) of the stent. Moreover, both the R-phase and martensite phase bands are coincident. This may be attributed to the R-phase acting as an intermediary phase in the B2 $\rightarrow$ B19' transformation, and therefore, both R-phase and B19' phases are forming cooperatively.

\begin{figure*}[h!]
    \includegraphics[width=180mm]{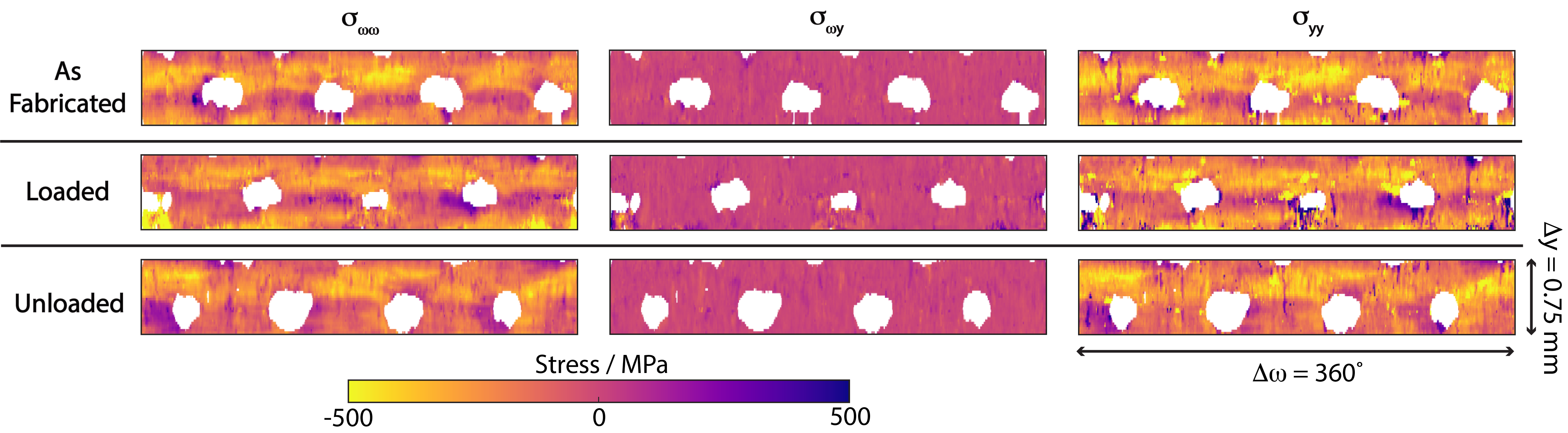}
    \caption{Stress tensor maps for the as-fabricated, deformed, and unloaded state of the stents.}
    \label{fig:Stress}
\end{figure*}

From the measured lattice strains for the B2 austenite phase, the macromechanical stresses were calculated as a function of stent location, as shown in Fig. \ref{fig:Stress}. A wide range of stresses ($\pm500$\,MPa), in all deformation states, was evident in the principal directions ($\omega$ and $y$), with near zero shear stress components. There is a strong correlation between regions of tensile or compressive stress for the $\sigma_{\omega\omega}$ and $\sigma_{yy}$ tensor components.  Given the geometry follows cylindrical symmetry, this result is unsurprising. Higher stresses were accumulated near junctions (locations between the empty spaces), with these regions propagating in the $\omega$ direction; this is far more evident in the deformed compared to the as fabricated state. The shape and distribution of the stress fields is evidently different following loading, indicating the magnitudes and their locations is a function of both the external stress and the stent geometry. Following unloading, the stress distribution is again modified, indicating residual stresses prevail that have been influenced by the material deformation mechanisms.

\section{Discussion}

\subsection{Transformation temperatures}
The broad transformation peaks were observed for the additively manufactured stent, as shown in Fig. \ref{fig:DSC}. These broad peaks may be attributed to the very localized heterogeneity, residual stresses, and additive manufacturing-induced defects. It is well known that a heat treatment will remote these artefacts and can be used to customise the transformation temperature \cite{Sam2018}. Similar broad peaks were reported by several authors for the additively manufactured specimens. For near equiatomic NiTi, the high-temperature austenite phase (B2) has two reported possible transformation paths; either into a martensitic phase (B19') or an R-phase during cooling \cite{Otsuka2005}. Figure \ref{fig:DSC} depicts the temperature hysteresis of $<10^\circ$C, indicating the B2 $\rightarrow$ R phase transformation during cooling (usually the hysteresis shall be $<10^\circ$C for B2$\rightarrow$R transformation) \cite{Vashishtha2022}. 

\subsection{Post-mortem analysis of stent}

The average hardness of $7.28\pm2.25$\,GPa is high, suggesting that the powder material fused properly, and good quality stents have been produced. As reported, the hardness was location specific, with preferred direction evident. To rationalise this, EBSD scans were recorded as shown in Fig. \ref{fig:EBSD}.  The IPF-$Z$ map in Fig. \ref{fig:EBSD} (b) shows elongated grains in the building direction. During the LPBF process and the subsequent epitaxial solidification, the grain growth follows the building direction as it possesses the highest heat flow rate \cite{Safdel2023}. The high level of residual stresses in the as-fabricated conditions, measured via diffraction, supports role of thermal effects during stent manufacture. Further, the EBSD scan at the junction of the ligaments shows a complex combination of fine grains outwards and elongated grains inwards, as shown in Fig. \ref{fig:EBSD} (e). This is attributed to the faster cooling rate and higher grain nucleation probability \cite{Akram2018} at the outward edge or ligament junction. Moreover, the LAGBs in the junction regions ($\sim$21\%) are higher than in the bulk region ($\sim$16.7\%) of the ligament. Therefore, it is evident that the higher LAGBs (or regions of higher dislocation density) and the fine grains correlate to the high hardness. The directionality of the measured hardness (Fig. \ref{fig:Nanoindentation}) is deemed a geometric effect, which likely connects edge locations where stress concentrations are highest. Thus, geometry has an important role in determining the microstructures and subsequently the mechanical properties. This must affect any subsequent process that makes such stents suitable in their intended applications.

\subsection{Phase formations, stress mapping, and tomographic reconstruction}

The phase formation in the as-received, deformed, and unloaded stents were detected by isolating the intensities of the diffraction patterns in Fig. \ref{fig:XRD}, particularly in the $B2_{(110)}$ austenitic phase, $B19'_{(200)}$ martensitic phase, and $R_{(122)}$ phase. Location evidently plays a role in phase formation; the stress concentration has been increased at the edges of struts in the deformed state. The B19' martensite phase that existed in the as-received state has been intensified in the deformation state, especially at the high-stress concentration edges, as shown in Fig. \ref{fig:Intensities}.  This may be attributed to the stent geometry at ligament junctions; these experiencing higher stresses and exceeds the critical stress to trigger the stress-induced transformation. Moreover, the R-phase (an intermediary phase in B2 $\rightarrow$ B19' transformation) particularly emerged in the deformed state of the stent and disappeared after load removal. The R-phase formation may be attributed to the inhomogeneity in the matrix \cite{Wang2016}. Moreover, the inhomogeneities in additively manufactured components may well promote the B2 $\rightarrow$ R transformation. \v{S}ittner et al. \cite{Sittner2006} also reported the formation of R-phase during tensile and compression loading of NiTi shape memory alloy by the neutron diffraction method.

\begin{figure}[h!]
 \centering
    \includegraphics[width=85mm]{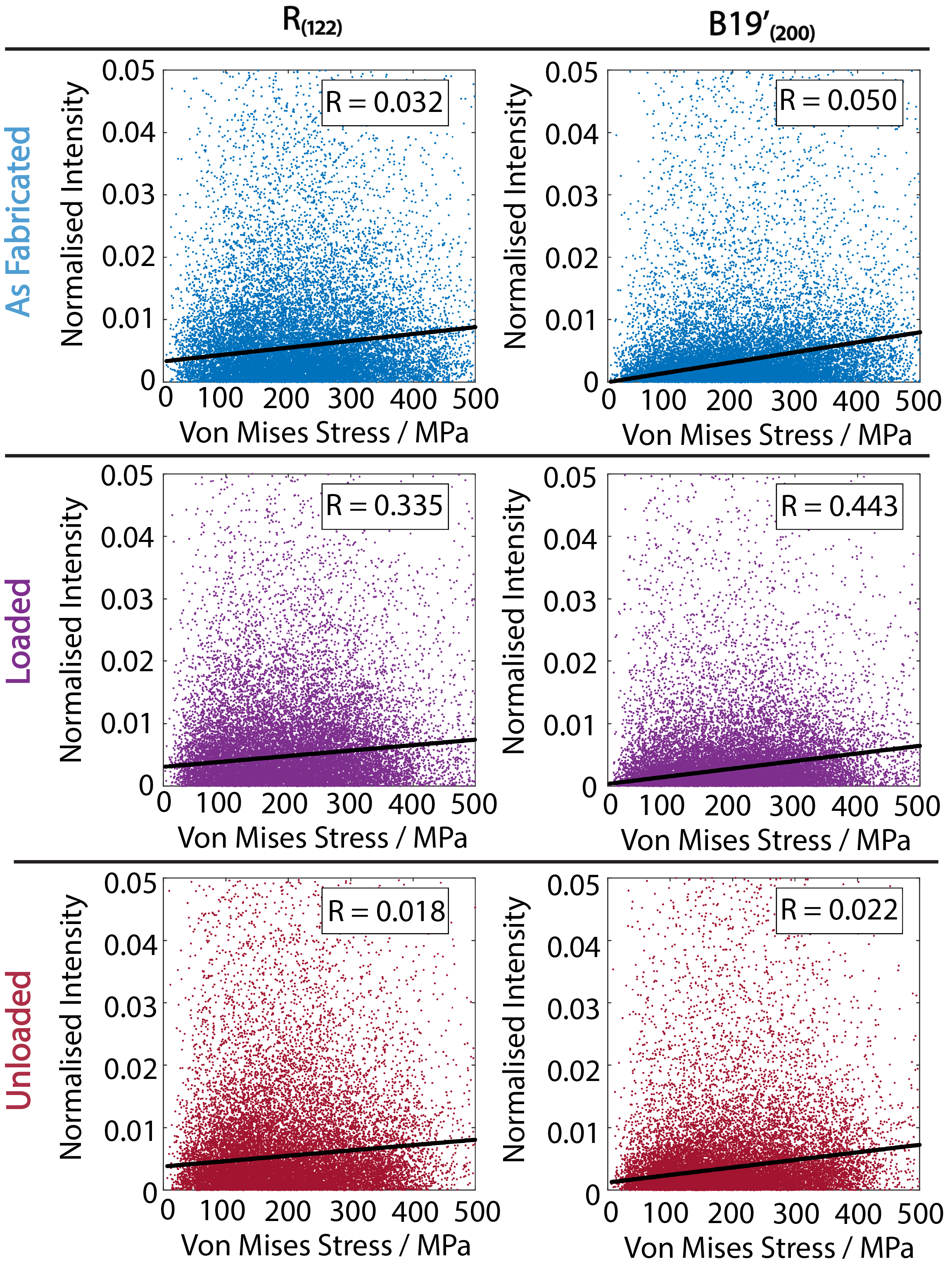}
    \caption{Correlation between the Von Mises Stress in the as fabricated, loaded and unloaded conditions with the intensities for the (122) reflection of the R phase and the (200) reflection of the martensitic B19' phase. Each coloured data point corresponds to a single diffraction pattern of the stent. The black lines are first order polynomials fitted to each dataset.}
    \label{fig:StressCorr}
\end{figure}

It can be seen qualitatively from Fig. \ref{fig:Intensities} and \ref{fig:Stress} that a strong correlation exists between locations of high stress and the presence of martensite/R phase. Given heterogeneous stress distribution, influenced by geometry, and locations that do or do not experience phase transformations, there are significant implications to the use application. To further corrborate this finding, the von Mises stress was plotted against normalised intensity with respect to each data point for $R_{(1\,2\,2)}$ and $B19'_{(2\,0\,0)}$ phases, as shown in Fig. \ref{fig:StressCorr}. To assess the correlation between these parameters, a linear regression fit was applied to the as-fabricated, loaded and unloaded state datasets. The correlation coefficient (R) is close to zero in as-fabricated and unloaded conditions for both $R_{(122)}$ and $B19'_{(2\,0\,0)}$ phases. However, the correlation coefficient increases in the loaded condition for both $R_{(1\,2\,2)}$ and $B19'_{(2\,0\,0)}$ phases. Therefore, it can be stated that the $R_{(1\,2\,2)}$ and $B19'_{(2\,0\,0)}$ phases are promoted by locations of increased localised stress. Whilst the out-of-plane stress components have been neglected in this 2D calculation of von Mises stress, so are likely to underestimate absolute values, this treatment serves as a suitable qualitative approximation for transformation-location correlations.

The implications of localised stress induced transformations have clear implications to the use of nitinol in cardiovascular stent applications. The results of this study indicate that regions experiencing a stress-induced phase transformation must also undergo cyclic damage \cite{Robertson2013}, and thereby exhaust their ability to undergo further austenite $\leftrightarrow$ phase transformations. Consequently, stent components will ensure location dependent cyclic amnesia, which will be component life limiting \cite{Abad2012}. Therefore, it is clear that ligament junctions experience the highest localised stresses, and are most susceptible to failure/stress-induced transformations. This highlights the requirement for careful geometry control when designing the stents. Whilst this may seem a pessimistic remark for the application of NiTi for stent applications, control of the fabrication process, microstructures \& residual stresses (for example), in concordance with geometric effects, can be exploited for acceptable component performance. 

\section{Conclusion}
In this investigation, LPBF fabricated shape memory alloy stents, of near equiatomic NiTi, were examined under in-situ loading and unloading conditions, combining X-ray imaging and diffraction. Specially designed miniaturised stents were created to perform a first-of-its-type study at the DIAD beamline, Diamond Light Source. These spatially resolved measurements reveal that geometry-material interactions determine localised deformation. The specific conclusions can be drawn are:
\begin{enumerate}

\item When a cylindrical strain was applied to the stent ($\varepsilon \approx1\%$), the stress-induced B19' martensite and R phase were formed from the parent B2 structured austenite phase. The stress-induced phases were localised, and were found together, indicating their concomitant formation. Removing the applied stress sees some but not all of the austenite phase reforming. 

\item The macromechanical stresses of the austenite phase, measured to a $25\times25\,\upmu$m$^2$ resolution, reveal linear stresses dominate over shear stresses, with the greatest magnitudes ($\pm 400$\,MPa) occurring between stent junctions, both in the loaded and unloaded states. These regions act as the sites most likely to form the martensite \& R phases. A moderate positive correlation is evident between the von Mises stress state and the R-phase \& B19' phases, confirming its role in their formation.

\item EBSD characterisation of the as fabricated state shows finer grains with greater lattice distortion in regions near stent ligament junctions, features inherited from the LPBF process. As these regions act as stress concentrations during loading, this explains the localised and highly heterogeneous phase distribution.

\item Nanoindentation mapping of a region near a stent junction adds credence to the heterogeneous deformation patterning. Factors such as porosity, directionality in the microstructure and thermally induced residual stress, varying with position, undoubtedly determine the locations most likely to promote stress induced phase transformations. 

\item Regions that transform preferentially when a stress is applied will be susceptible to degradation processes such as cyclic amnesia, well known for NiTi shape memory alloys, and would determine failure initiation. Such post-processing of LPBF stents must control this behaviour for part longevity for in-vivo applications.

\end{enumerate}

\section{Acknowledgments}

DMC and HV acknowledge the funding from a Leverhulme Trust Research Project Grant (RPG-2021-244). MMA acknowledges the EPSRC (EP/R00160X/1) for their financial contributions.  The authors are also grateful to the Diamond Light Source (Experiment Number MG28029-1) for the provision of beamtime and for access to laboratory space.  The authors are also thankful to Rebecca Garrard for her help in fabricating the stents.


\begin{thebibliography}{58}
\expandafter\ifx\csname natexlab\endcsname\relax\def\natexlab#1{#1}\fi
\providecommand{\url}[1]{\texttt{#1}}
\providecommand{\href}[2]{#2}
\providecommand{\path}[1]{#1}
\providecommand{\DOIprefix}{doi:}
\providecommand{\ArXivprefix}{arXiv:}
\providecommand{\URLprefix}{URL: }
\providecommand{\Pubmedprefix}{pmid:}
\providecommand{\doi}[1]{\href{http://dx.doi.org/#1}{\path{#1}}}
\providecommand{\Pubmed}[1]{\href{pmid:#1}{\path{#1}}}
\providecommand{\bibinfo}[2]{#2}
\ifx\xfnm\relax \def\xfnm[#1]{\unskip,\space#1}\fi
\bibitem[{Walke et~al.(2005)Walke, Paszenda, and Filipiak}]{Walke2005}
\bibinfo{author}{W.~Walke}, \bibinfo{author}{Z.~Paszenda},
  \bibinfo{author}{J.~Filipiak},
\newblock \bibinfo{journal}{J. Mater. Process. Technol.}
  \bibinfo{volume}{164–165} (\bibinfo{year}{2005})
  \bibinfo{pages}{1263--1268}.
\bibitem[{Kapnisis et~al.(2014)Kapnisis, Constantinides, Georgiou, Cristea,
  Gabor, Munteanu, Brott, Anderson, Lemons, and Anayiotos}]{Kapnisis2014}
\bibinfo{author}{K.~Kapnisis}, \bibinfo{author}{G.~Constantinides},
  \bibinfo{author}{H.~Georgiou}, \bibinfo{author}{D.~Cristea},
  \bibinfo{author}{C.~Gabor}, \bibinfo{author}{D.~Munteanu},
  \bibinfo{author}{B.~Brott}, \bibinfo{author}{P.~Anderson},
  \bibinfo{author}{J.~Lemons}, \bibinfo{author}{A.~Anayiotos},
\newblock \bibinfo{journal}{J. Mech. Behav. Biomed. Mater.}
  \bibinfo{volume}{40} (\bibinfo{year}{2014}) \bibinfo{pages}{240--251}.
\bibitem[{Sweeney et~al.(2014)Sweeney, O'Brien, McHugh, and Leen}]{Sweeney2014}
\bibinfo{author}{C.~A. Sweeney}, \bibinfo{author}{B.~O'Brien},
  \bibinfo{author}{P.~E. McHugh}, \bibinfo{author}{S.~B. Leen},
\newblock \bibinfo{journal}{Biomaterials} \bibinfo{volume}{35}
  (\bibinfo{year}{2014}) \bibinfo{pages}{36–48}.
\bibitem[{Sweeney et~al.(2015)Sweeney, O'Brien, Dunne, McHugh, and
  Leen}]{Sweeney2015}
\bibinfo{author}{C.~A. Sweeney}, \bibinfo{author}{B.~O'Brien},
  \bibinfo{author}{F.~P.~E. Dunne}, \bibinfo{author}{P.~E. McHugh},
  \bibinfo{author}{S.~B. Leen},
\newblock \bibinfo{journal}{J. Mech. Behav. Biomed. Mater.}
  \bibinfo{volume}{46} (\bibinfo{year}{2015}) \bibinfo{pages}{244–260}.
\bibitem[{Grogan et~al.(2014)Grogan, Leen, and McHugh}]{Grogan2014}
\bibinfo{author}{J.~A. Grogan}, \bibinfo{author}{S.~B. Leen},
  \bibinfo{author}{P.~E. McHugh},
\newblock \bibinfo{journal}{J. Mech. Behav. Biomed. Mater.}
  \bibinfo{volume}{34} (\bibinfo{year}{2014}) \bibinfo{pages}{93–105}.
\bibitem[{Amani et~al.(2017)Amani, Faraji, Mehrabadi, Abrinia, and
  Ghanbari}]{Amani2017}
\bibinfo{author}{S.~Amani}, \bibinfo{author}{G.~Faraji}, \bibinfo{author}{H.~K.
  Mehrabadi}, \bibinfo{author}{K.~Abrinia}, \bibinfo{author}{H.~Ghanbari},
\newblock \bibinfo{journal}{J. Alloys Compd.} \bibinfo{volume}{723}
  (\bibinfo{year}{2017}) \bibinfo{pages}{467–476}.
\bibitem[{Hsiao et~al.(2014)Hsiao, Wu, Yin, Lin, and Chen}]{Hsiao2014}
\bibinfo{author}{H.~M. Hsiao}, \bibinfo{author}{L.~W. Wu},
  \bibinfo{author}{M.~T. Yin}, \bibinfo{author}{C.~H. Lin},
  \bibinfo{author}{H.~Chen},
\newblock \bibinfo{journal}{Comput. Mater. Sci.} \bibinfo{volume}{86}
  (\bibinfo{year}{2014}) \bibinfo{pages}{57–63}.
\bibitem[{McGrath et~al.(2014)McGrath, O'Brien, Bruzzi, and
  McHugh}]{McGrath2014}
\bibinfo{author}{D.~J. McGrath}, \bibinfo{author}{B.~O'Brien},
  \bibinfo{author}{M.~Bruzzi}, \bibinfo{author}{P.~E. McHugh},
\newblock \bibinfo{journal}{J. Mech. Behav. Biomed. Mater.}
  \bibinfo{volume}{40} (\bibinfo{year}{2014}) \bibinfo{pages}{252–263}.
\bibitem[{Feng et~al.(2020)Feng, Kong, Hao, Liu, Yang, Yang, Guo, Jiang, Wang,
  Ren, and Cui}]{Feng2020}
\bibinfo{author}{B.~Feng}, \bibinfo{author}{X.~Kong}, \bibinfo{author}{S.~Hao},
  \bibinfo{author}{Y.~Liu}, \bibinfo{author}{Y.~Yang},
  \bibinfo{author}{H.~Yang}, \bibinfo{author}{F.~Guo},
  \bibinfo{author}{D.~Jiang}, \bibinfo{author}{T.~Wang},
  \bibinfo{author}{Y.~Ren}, \bibinfo{author}{L.~Cui},
\newblock \bibinfo{journal}{Acta Mater.} \bibinfo{volume}{194}
  (\bibinfo{year}{2020}) \bibinfo{pages}{565–576}.
\bibitem[{Duerig and Bhattacharya(2015)}]{Duerig2015}
\bibinfo{author}{T.~W. Duerig}, \bibinfo{author}{K.~Bhattacharya},
\newblock \bibinfo{journal}{Shape Mem. Superelasticity.} \bibinfo{volume}{1}
  (\bibinfo{year}{2015}) \bibinfo{pages}{153--161}.
\bibitem[{Zhao et~al.(2023)Zhao, Yu, Ren, Wang, Zhang, Chen, Xu, Ren, and
  Qu}]{Zhao2023}
\bibinfo{author}{Y.~Zhao}, \bibinfo{author}{Z.~Yu}, \bibinfo{author}{X.~Ren},
  \bibinfo{author}{Q.~Wang}, \bibinfo{author}{B.~Zhang},
  \bibinfo{author}{J.~Chen}, \bibinfo{author}{W.~Xu}, \bibinfo{author}{S.~Ren},
  \bibinfo{author}{X.~Qu},
\newblock \bibinfo{journal}{Int. J. Fatigue} \bibinfo{volume}{169}
  (\bibinfo{year}{2023}) \bibinfo{pages}{107494}.
\bibitem[{Korei et~al.(2022)Korei, Solouk, Nazarpak, and Nouri}]{Korei2022a}
\bibinfo{author}{N.~Korei}, \bibinfo{author}{A.~Solouk},
  \bibinfo{author}{H.~Nazarpak}, \bibinfo{author}{A.~Nouri},
\newblock \bibinfo{journal}{Mater. Today Commun.} \bibinfo{volume}{31}
  (\bibinfo{year}{2022}) \bibinfo{pages}{103467}.
\bibitem[{Dong et~al.(2018)Dong, Li, Wang, and Zhou}]{Dong2018}
\bibinfo{author}{P.~Dong}, \bibinfo{author}{H.~Li}, \bibinfo{author}{W.~Wang},
  \bibinfo{author}{J.~Zhou},
\newblock \bibinfo{journal}{Mater. Charact.} \bibinfo{volume}{135}
  (\bibinfo{year}{2018}) \bibinfo{pages}{40–45}.
\bibitem[{Biffi et~al.(2022)Biffi, Fiocchi, and Tuissi}]{Biffi2022}
\bibinfo{author}{C.~A. Biffi}, \bibinfo{author}{J.~Fiocchi},
  \bibinfo{author}{A.~Tuissi},
\newblock \bibinfo{journal}{J. Mater. Res. Technol.} \bibinfo{volume}{19}
  (\bibinfo{year}{2022}) \bibinfo{pages}{472–506}.
\bibitem[{Fu et~al.(2014)Fu, Guo, and Sealy}]{Fu2014}
\bibinfo{author}{C.~H. Fu}, \bibinfo{author}{Y.~B. Guo}, \bibinfo{author}{M.~P.
  Sealy},
\newblock \bibinfo{journal}{J. Mater. Process. Technol.} \bibinfo{volume}{214}
  (\bibinfo{year}{2014}) \bibinfo{pages}{2926–2934}.
\bibitem[{Cheng et~al.(2022)Cheng, Zhang, Liu, Li, Zeng, Zhou, and
  Zhao}]{Cheng2022}
\bibinfo{author}{Y.~Cheng}, \bibinfo{author}{X.~Zhang},
  \bibinfo{author}{R.~Liu}, \bibinfo{author}{Y.~Li}, \bibinfo{author}{J.~Zeng},
  \bibinfo{author}{M.~Zhou}, \bibinfo{author}{Y.~Zhao},
\newblock \bibinfo{journal}{Adv. Healthc. Mater.} \bibinfo{volume}{11}
  (\bibinfo{year}{2022}) \bibinfo{pages}{2200965}.
\bibitem[{Lee et~al.(2023)Lee, In, Park, and Kim}]{Lee2023}
\bibinfo{author}{J.~C. Lee}, \bibinfo{author}{S.~H. In}, \bibinfo{author}{C.~H.
  Park}, \bibinfo{author}{C.~S. Kim},
\newblock \bibinfo{journal}{Mater. Lett.} \bibinfo{volume}{331}
  (\bibinfo{year}{2023}) \bibinfo{pages}{133415}.
\bibitem[{Li et~al.(2023)Li, Shi, Lu, Li, Zhou, Zadpoor, and Wang}]{Li2023}
\bibinfo{author}{Y.~Li}, \bibinfo{author}{Y.~Shi}, \bibinfo{author}{Y.~Lu},
  \bibinfo{author}{X.~Li}, \bibinfo{author}{J.~Zhou}, \bibinfo{author}{A.~A.
  Zadpoor}, \bibinfo{author}{L.~Wang},
\newblock \bibinfo{journal}{Acta Biomater.} \bibinfo{volume}{167}
  (\bibinfo{year}{2023}) \bibinfo{pages}{16–37}.
\bibitem[{Wiesent et~al.(2023)Wiesent, Stocker, and Nonn}]{Wiesent2023}
\bibinfo{author}{L.~Wiesent}, \bibinfo{author}{F.~Stocker},
  \bibinfo{author}{A.~Nonn},
\newblock \bibinfo{journal}{Materialia} \bibinfo{volume}{28}
  (\bibinfo{year}{2023}) \bibinfo{pages}{101774}.
\bibitem[{Safdel and Elbestawi(2021)}]{Safdel2021}
\bibinfo{author}{A.~Safdel}, \bibinfo{author}{M.~A. Elbestawi},
\newblock \bibinfo{journal}{Mater. Lett.} \bibinfo{volume}{300}
  (\bibinfo{year}{2021}) \bibinfo{pages}{130163}.
\bibitem[{Fu et~al.(2015)Fu, Liu, and Guo}]{Fu2015}
\bibinfo{author}{C.~H. Fu}, \bibinfo{author}{J.~F. Liu},
  \bibinfo{author}{A.~Guo},
\newblock \bibinfo{journal}{Appl. Surf. Sci.} \bibinfo{volume}{353}
  (\bibinfo{year}{2015}) \bibinfo{pages}{291–299}.
\bibitem[{Pan et~al.(2021)Pan, Han, and Lu}]{Pan2021}
\bibinfo{author}{C.~Pan}, \bibinfo{author}{Y.~Han}, \bibinfo{author}{J.~Lu},
\newblock \bibinfo{journal}{Micromachines} \bibinfo{volume}{12}
  (\bibinfo{year}{2021}) \bibinfo{pages}{770}.
\bibitem[{Korei et~al.(2022)Korei, Solouk, Nazarpak, and Nouri}]{Korei2022b}
\bibinfo{author}{N.~Korei}, \bibinfo{author}{A.~Solouk},
  \bibinfo{author}{H.~Nazarpak}, \bibinfo{author}{A.~Nouri},
\newblock \bibinfo{journal}{Mater. Today Commun.} \bibinfo{volume}{31}
  (\bibinfo{year}{2022}) \bibinfo{pages}{103467}.
\bibitem[{Chowdhury et~al.(2022)Chowdhury, Yadaiah, Prakash, Ramakrishna,
  Dixit, Gupta, and Buddhi}]{Chowdhury2022}
\bibinfo{author}{S.~Chowdhury}, \bibinfo{author}{N.~Yadaiah},
  \bibinfo{author}{C.~Prakash}, \bibinfo{author}{S.~Ramakrishna},
  \bibinfo{author}{S.~Dixit}, \bibinfo{author}{L.~Gupta},
  \bibinfo{author}{D.~Buddhi},
\newblock \bibinfo{journal}{J. Mater. Res. Technol.} \bibinfo{volume}{20}
  (\bibinfo{year}{2022}) \bibinfo{pages}{2109--2172}.
\bibitem[{Stoeckel et~al.(2002)Stoeckel, Bonsignore, and Duda}]{Stoeckel2002}
\bibinfo{author}{D.~Stoeckel}, \bibinfo{author}{C.~Bonsignore},
  \bibinfo{author}{S.~Duda},
\newblock \bibinfo{journal}{Minim. Invasive Ther. Allied Technol.}
  \bibinfo{volume}{11} (\bibinfo{year}{2002}) \bibinfo{pages}{137–147}.
\bibitem[{Ahadi et~al.(2023)Ahadi, Azadi, Biglari, Bodaghi, and
  Khaleghian}]{Ahadi2023}
\bibinfo{author}{F.~Ahadi}, \bibinfo{author}{M.~Azadi},
  \bibinfo{author}{M.~Biglari}, \bibinfo{author}{M.~Bodaghi},
  \bibinfo{author}{A.~Khaleghian},
\newblock \bibinfo{journal}{Heliyon} \bibinfo{volume}{9} (\bibinfo{year}{2023})
  \bibinfo{pages}{e13575}.
\bibitem[{Wiesent et~al.(2022)Wiesent, Spear, and Nonn}]{Wiesent2022}
\bibinfo{author}{L.~Wiesent}, \bibinfo{author}{A.~Spear},
  \bibinfo{author}{A.~Nonn},
\newblock \bibinfo{journal}{J. Mech. Behav. Biomed. Mater.}
  \bibinfo{volume}{125} (\bibinfo{year}{2022}) \bibinfo{pages}{104878}.
\bibitem[{Berti et~al.(2021)Berti, Wang, Spagnoli, Pennati, Migliavacca,
  Edelman, and Petrini}]{Berti2021}
\bibinfo{author}{F.~Berti}, \bibinfo{author}{P.~J. Wang},
  \bibinfo{author}{A.~Spagnoli}, \bibinfo{author}{G.~Pennati},
  \bibinfo{author}{F.~Migliavacca}, \bibinfo{author}{E.~R. Edelman},
  \bibinfo{author}{L.~Petrini},
\newblock \bibinfo{journal}{J. Mech. Behav. Biomed. Mater.}
  \bibinfo{volume}{113} (\bibinfo{year}{2021}) \bibinfo{pages}{104142}.
\bibitem[{Berti et~al.(2019)Berti, Spagnoli, and Petrini}]{Berti2019}
\bibinfo{author}{F.~Berti}, \bibinfo{author}{A.~Spagnoli},
  \bibinfo{author}{L.~Petrini},
\newblock \bibinfo{journal}{Eng. Fract. Mech.} \bibinfo{volume}{216}
  (\bibinfo{year}{2019}) \bibinfo{pages}{106512}.
\bibitem[{Jedwab and Clerc(1993)}]{Jedwab1993}
\bibinfo{author}{M.~R. Jedwab}, \bibinfo{author}{C.~O. Clerc},
\newblock \bibinfo{journal}{J. Appl. Biomater.} \bibinfo{volume}{4}
  (\bibinfo{year}{1993}) \bibinfo{pages}{77–85}.
\bibitem[{Raghunathan et~al.(2008)Raghunathan, Azeem, Collins, and
  Dye}]{RAGHUNATHAN20081059}
\bibinfo{author}{S.~Raghunathan}, \bibinfo{author}{M.~Azeem},
  \bibinfo{author}{D.~Collins}, \bibinfo{author}{D.~Dye},
\newblock \bibinfo{journal}{Scripta Mater.} \bibinfo{volume}{59}
  (\bibinfo{year}{2008}) \bibinfo{pages}{1059--1062}.
\bibitem[{Safdel et~al.(2023)Safdel, Torbati-Sarraf, and
  Elbestawi}]{Safdel2023}
\bibinfo{author}{A.~Safdel}, \bibinfo{author}{H.~Torbati-Sarraf},
  \bibinfo{author}{M.~A. Elbestawi},
\newblock \bibinfo{journal}{J. Alloys Compd.} \bibinfo{volume}{954}
  (\bibinfo{year}{2023}) \bibinfo{pages}{170196}.
\bibitem[{Jamshidi et~al.(2022)Jamshidi, Panwisawas, Langi, Cox, Feng, Zhao,
  and Attallah}]{Jamshidi2022}
\bibinfo{author}{P.~Jamshidi}, \bibinfo{author}{C.~Panwisawas},
  \bibinfo{author}{E.~Langi}, \bibinfo{author}{S.~C. Cox},
  \bibinfo{author}{J.~Feng}, \bibinfo{author}{L.~Zhao}, \bibinfo{author}{M.~M.
  Attallah},
\newblock \bibinfo{journal}{J. Alloys Compd.} \bibinfo{volume}{909}
  (\bibinfo{year}{2022}) \bibinfo{pages}{164681}.
\bibitem[{Bian et~al.(2022)Bian, Heller, Kadeř\'{a}vek, and
  \v{S}ittner}]{Bian2022}
\bibinfo{author}{X.~Bian}, \bibinfo{author}{L.~Heller},
  \bibinfo{author}{L.~Kadeř\'{a}vek}, \bibinfo{author}{P.~\v{S}ittner},
\newblock \bibinfo{journal}{Appl. Mater. Today} \bibinfo{volume}{26}
  (\bibinfo{year}{2022}) \bibinfo{pages}{101378}.
\bibitem[{Sedm\'{a}k et~al.(2015)Sedm\'{a}k, \v{S}ittner, Pilch, and
  Curfs}]{Sedmak2015}
\bibinfo{author}{P.~Sedm\'{a}k}, \bibinfo{author}{P.~\v{S}ittner},
  \bibinfo{author}{J.~Pilch}, \bibinfo{author}{C.~Curfs},
\newblock \bibinfo{journal}{Acta Mater.} \bibinfo{volume}{94}
  (\bibinfo{year}{2015}) \bibinfo{pages}{257–270}.
\bibitem[{Urbina et~al.(2017)Urbina, Gispert-Guirado, Ferrando, and
  la~Flor}]{Urbina2017}
\bibinfo{author}{C.~Urbina}, \bibinfo{author}{F.~Gispert-Guirado},
  \bibinfo{author}{F.~Ferrando}, \bibinfo{author}{S.~D. la~Flor},
\newblock \bibinfo{journal}{J. Alloys Compd.} \bibinfo{volume}{712}
  (\bibinfo{year}{2017}) \bibinfo{pages}{833–847}.
\bibitem[{Yu et~al.(2016)Yu, Aoun, Cui, Liu, Yang, Jiang, Cai, Jiang, Liu,
  Brown, and Ren}]{Yu2016}
\bibinfo{author}{C.~Yu}, \bibinfo{author}{B.~Aoun}, \bibinfo{author}{L.~Cui},
  \bibinfo{author}{Y.~Liu}, \bibinfo{author}{H.~Yang},
  \bibinfo{author}{X.~Jiang}, \bibinfo{author}{S.~Cai},
  \bibinfo{author}{D.~Jiang}, \bibinfo{author}{Z.~Liu}, \bibinfo{author}{D.~E.
  Brown}, \bibinfo{author}{Y.~Ren},
\newblock \bibinfo{journal}{Acta Mater.} \bibinfo{volume}{115}
  (\bibinfo{year}{2016}) \bibinfo{pages}{35–44}.
\bibitem[{Reinhard et~al.(2021)Reinhard, Drakopoulos, Ahmed, Deyhle, James,
  Charlesworth, Burt, Sutter, Alexander, Garland, Yates, Marshall, Kemp,
  Warrick, Pueyos, Bradnick, Nagni, Winter, Filik, Basham, Wadeson, King,
  Aslani, and Dent}]{Reinhard2021}
\bibinfo{author}{C.~Reinhard}, \bibinfo{author}{M.~Drakopoulos},
  \bibinfo{author}{S.~I. Ahmed}, \bibinfo{author}{H.~Deyhle},
  \bibinfo{author}{A.~James}, \bibinfo{author}{C.~M. Charlesworth},
  \bibinfo{author}{M.~Burt}, \bibinfo{author}{J.~Sutter},
  \bibinfo{author}{S.~Alexander}, \bibinfo{author}{P.~Garland},
  \bibinfo{author}{T.~Yates}, \bibinfo{author}{R.~Marshall},
  \bibinfo{author}{B.~Kemp}, \bibinfo{author}{E.~Warrick},
  \bibinfo{author}{A.~Pueyos}, \bibinfo{author}{B.~Bradnick},
  \bibinfo{author}{M.~Nagni}, \bibinfo{author}{A.~D. Winter},
  \bibinfo{author}{J.~Filik}, \bibinfo{author}{M.~Basham},
  \bibinfo{author}{N.~Wadeson}, \bibinfo{author}{O.~N.~F. King},
  \bibinfo{author}{N.~Aslani}, \bibinfo{author}{A.~J. Dent},
\newblock \bibinfo{journal}{J. Synchrotron Radiat.} \bibinfo{volume}{28}
  (\bibinfo{year}{2021}) \bibinfo{pages}{1985–1995}.
\bibitem[{Saedi et~al.(2018)Saedi, Moghaddam, Amerinatanzi, Elahinia, and
  Karaca}]{Saedi2018}
\bibinfo{author}{S.~Saedi}, \bibinfo{author}{S.~Moghaddam},
  \bibinfo{author}{A.~Amerinatanzi}, \bibinfo{author}{M.~Elahinia},
  \bibinfo{author}{H.~E. Karaca},
\newblock \bibinfo{journal}{Acta Mater.} \bibinfo{volume}{144}
  (\bibinfo{year}{2018}) \bibinfo{pages}{552--560}.
\bibitem[{Vashishtha and Jain(2022)}]{Vashishtha2022}
\bibinfo{author}{H.~Vashishtha}, \bibinfo{author}{J.~Jain},
\newblock \bibinfo{journal}{J. Alloys Compd.} \bibinfo{volume}{893}
  (\bibinfo{year}{2022}) \bibinfo{pages}{162307}.
\bibitem[{Otsuka and Ren(2005)}]{Otsuka2005}
\bibinfo{author}{K.~Otsuka}, \bibinfo{author}{X.~Ren},
\newblock \bibinfo{journal}{Prog. Mater. Sci.} \bibinfo{volume}{50}
  (\bibinfo{year}{2005}) \bibinfo{pages}{511–678}.
\bibitem[{Chen et~al.(2021)Chen, Clark, Collins, Marussi, Hunt, Fenech,
  Connolley, Atwood, Magdysyuk, Baxter, Jones, Leung, and Lee}]{CHEN2021116777}
\bibinfo{author}{Y.~Chen}, \bibinfo{author}{S.~J. Clark},
  \bibinfo{author}{D.~M. Collins}, \bibinfo{author}{S.~Marussi},
  \bibinfo{author}{S.~A. Hunt}, \bibinfo{author}{D.~M. Fenech},
  \bibinfo{author}{T.~Connolley}, \bibinfo{author}{R.~C. Atwood},
  \bibinfo{author}{O.~V. Magdysyuk}, \bibinfo{author}{G.~J. Baxter},
  \bibinfo{author}{M.~A. Jones}, \bibinfo{author}{C.~L.~A. Leung},
  \bibinfo{author}{P.~D. Lee},
\newblock \bibinfo{journal}{Acta Mater.} \bibinfo{volume}{209}
  (\bibinfo{year}{2021}) \bibinfo{pages}{116777}.
\bibitem[{He(2011)}]{He}
\bibinfo{author}{B.~He}, \bibinfo{title}{Two-dimensional X-ray diffraction},
  \bibinfo{address}{Hoboken, NJ}, \bibinfo{year}{2011}.
\bibitem[{Hauk(1997)}]{Hauk}
\bibinfo{author}{V.~Hauk}, \bibinfo{title}{Structural and Residual Stress
  Analysis by Nondestructive Methods}, \bibinfo{publisher}{Elsevier Science
  BV}, \bibinfo{address}{Amstertam}, \bibinfo{year}{1997}.
\bibitem[{Gn{\"{a}}upel-Herold(2012)}]{isodec}
\bibinfo{author}{T.~Gn{\"{a}}upel-Herold},
\newblock \bibinfo{journal}{J. Appl. Cryst.} \bibinfo{volume}{45}
  (\bibinfo{year}{2012}) \bibinfo{pages}{573--574}.
\bibitem[{Mercier et~al.(1980)Mercier, N.Melton, , and Gremaud}]{Mercier}
\bibinfo{author}{Q.~Mercier}, \bibinfo{author}{K.~N.Melton}, ,
  \bibinfo{author}{G.~Gremaud},
\newblock \bibinfo{journal}{J. Appl. Phys.} \bibinfo{volume}{51}
  (\bibinfo{year}{1980}) \bibinfo{pages}{1833--1837}.
\bibitem[{Welzel et~al.(2005)Welzel, Ligot, Lamparter, Vermeulen, and
  Mittemeijer}]{stress_review}
\bibinfo{author}{U.~Welzel}, \bibinfo{author}{J.~Ligot},
  \bibinfo{author}{P.~Lamparter}, \bibinfo{author}{A.~C. Vermeulen},
  \bibinfo{author}{E.~J. Mittemeijer} \bibinfo{volume}{38}
  (\bibinfo{year}{2005}) \bibinfo{pages}{1}.
\bibitem[{Suresh et~al.(2012)Suresh, Kim, Bhaumik, and Suwas}]{Suresh2012}
\bibinfo{author}{K.~S. Suresh}, \bibinfo{author}{D.-I. Kim},
  \bibinfo{author}{S.~K. Bhaumik}, \bibinfo{author}{S.~Suwas},
\newblock \bibinfo{journal}{Scr. Mater.} \bibinfo{volume}{66}
  (\bibinfo{year}{2012}) \bibinfo{pages}{602--605}.
\bibitem[{Pfetzing-Micklich et~al.(2013)Pfetzing-Micklich, Somsen, Dlouhy,
  Begau, Hartmaier, Wagner, and Eggeler}]{Pfetzing2013}
\bibinfo{author}{J.~Pfetzing-Micklich}, \bibinfo{author}{C.~Somsen},
  \bibinfo{author}{A.~Dlouhy}, \bibinfo{author}{C.~Begau},
  \bibinfo{author}{A.~Hartmaier}, \bibinfo{author}{M.~F.-X. Wagner},
  \bibinfo{author}{G.~Eggeler},
\newblock \bibinfo{journal}{Acta Mater.} \bibinfo{volume}{61}
  (\bibinfo{year}{2013}) \bibinfo{pages}{602--616}.
\bibitem[{du~Plessis et~al.(2020)du~Plessis, Yadroitsava, and
  Yadroitsev}]{Anton2020}
\bibinfo{author}{A.~du~Plessis}, \bibinfo{author}{I.~Yadroitsava},
  \bibinfo{author}{I.~Yadroitsev},
\newblock \bibinfo{journal}{Mater. Des.} \bibinfo{volume}{187}
  (\bibinfo{year}{2020}) \bibinfo{pages}{108385}.
\bibitem[{Kok et~al.(2018)Kok, Tan, Wang, Nai, Loh, Liu, and Tor}]{Kok2018}
\bibinfo{author}{Y.~Kok}, \bibinfo{author}{X.~P. Tan},
  \bibinfo{author}{P.~Wang}, \bibinfo{author}{M.~L.~S. Nai},
  \bibinfo{author}{N.~H. Loh}, \bibinfo{author}{E.~Liu}, \bibinfo{author}{S.~B.
  Tor},
\newblock \bibinfo{journal}{Mater. Des.} \bibinfo{volume}{139}
  (\bibinfo{year}{2018}) \bibinfo{pages}{565--586}.
\bibitem[{Snyder and Thole(2020)}]{Jacob2020}
\bibinfo{author}{J.~C. Snyder}, \bibinfo{author}{K.~A. Thole},
\newblock \bibinfo{journal}{J. Manuf. Sci. Eng.} \bibinfo{volume}{142}
  (\bibinfo{year}{2020}) \bibinfo{pages}{071003}.
\bibitem[{Sam et~al.(2018)Sam, Franco, Ma, Karaman, Elwany, and Mabe}]{Sam2018}
\bibinfo{author}{J.~Sam}, \bibinfo{author}{B.~Franco}, \bibinfo{author}{J.~Ma},
  \bibinfo{author}{I.~Karaman}, \bibinfo{author}{A.~Elwany},
  \bibinfo{author}{J.~H. Mabe},
\newblock \bibinfo{journal}{Scripta Mater.} \bibinfo{volume}{146}
  (\bibinfo{year}{2018}) \bibinfo{pages}{164--168}.
\bibitem[{Akram et~al.(2018)Akram, Chalavadi, Pal, and Stucker}]{Akram2018}
\bibinfo{author}{J.~Akram}, \bibinfo{author}{P.~Chalavadi},
  \bibinfo{author}{D.~Pal}, \bibinfo{author}{B.~Stucker},
\newblock \bibinfo{journal}{Addit. Manuf.} \bibinfo{volume}{21}
  (\bibinfo{year}{2018}) \bibinfo{pages}{255--268}.
\bibitem[{Wang et~al.(2016)Wang, Wang, Zhang, Chen, Lu, and Zhang}]{Wang2016}
\bibinfo{author}{L.~Wang}, \bibinfo{author}{C.~Wang},
  \bibinfo{author}{L.~Zhang}, \bibinfo{author}{L.~Chen},
  \bibinfo{author}{W.~Lu}, \bibinfo{author}{D.~Zhang},
\newblock \bibinfo{journal}{Sci. Rep.} \bibinfo{volume}{6}
  (\bibinfo{year}{2016}) \bibinfo{pages}{23905}.
\bibitem[{\v{S}ittner et~al.(2006)\v{S}ittner, Landa, Luk\'{a}\v{s}, and
  Nov\'{a}k}]{Sittner2006}
\bibinfo{author}{P.~\v{S}ittner}, \bibinfo{author}{M.~Landa},
  \bibinfo{author}{P.~Luk\'{a}\v{s}}, \bibinfo{author}{V.~Nov\'{a}k},
\newblock \bibinfo{journal}{Mech. Mater.} \bibinfo{volume}{38}
  (\bibinfo{year}{2006}) \bibinfo{pages}{475–492}.
\bibitem[{Robertson et~al.(2013)Robertson, Pelton, and Ritchie}]{Robertson2013}
\bibinfo{author}{S.~W. Robertson}, \bibinfo{author}{A.~R. Pelton},
  \bibinfo{author}{R.~O. Ritchie},
\newblock \bibinfo{journal}{Int. Mater. Rev.} \bibinfo{volume}{57}
  (\bibinfo{year}{2013}) \bibinfo{pages}{1--37}.
\bibitem[{Abad et~al.(2012)Abad, Pasini, and Cecere}]{Abad2012}
\bibinfo{author}{E.~M.~K. Abad}, \bibinfo{author}{D.~Pasini},
  \bibinfo{author}{R.~Cecere},
\newblock \bibinfo{journal}{J. Biomech.} \bibinfo{volume}{45}
  (\bibinfo{year}{2012}) \bibinfo{pages}{1028--1035}.

\end{thebibliography}
\bibliographystyle{elsarticle-num-names}

\end{document}